\pdfoutput=1
\documentclass[12pt]{article} 
\usepackage[margin=25mm]{geometry}

\usepackage{float}
\usepackage{tabularx,rotating}
\usepackage{multicol}
\usepackage{array, booktabs, makecell}
\usepackage{amsbsy,amsmath,amssymb}
\usepackage{graphicx}
\usepackage{color,soul}
\usepackage{authblk}
\usepackage[colorlinks,allcolors=blue]{hyperref}
\usepackage[capitalise]{cleveref}

\providecommand{\keywords}[1]
{
  \small	
  \textbf{\textit{Keywords---}} #1
}

\title{Importance of preferential segregation by aerodynamics in dust rig tests}
\author{Cairen Miranda, John Palmore Jr  \\
        \small Department of Mechanical Engineering,\\
		Virginia Tech,\\
		Blacksburg, Virginia, USA 24061 \\
}
\date{}

\begin{document}

	\maketitle 
	
	\begin{abstract}
		This work studies the design of a device conveying dust and sand in order to understand how the particles impinge, erode and rebound from a target plate. The motivation behind this study is understanding dust ingestion and erosion in aviation gas-turbine engines. The conveying system consists of a constant area duct in which particles are injected in the upstream direction using a particle seeder. The particles exit the duct through a converging nozzle and are directed towards a target plate. Particles of varying sizes are tested at two different gas speeds.
		
		The particle velocity at the conveyor duct exit follows a trend inversely proportional to diameter. After exiting, particles respond differently to changes in the flow field based on their diameter. The largest particles move ballistically, so they impact the target with nearly the same velocity they had at the duct exit. However, small particles follow the flow streamlines around the target. This causes them to both significantly slow down and to disperse in all directions. The combination of the exit velocity and the near-target trajectory behaviors leads to a non-monotonic trend of particle impact velocity as a function of particle diameter. The importance of these effects depends strongly on the relative angle between the conveyor duct and the target plate.
		
		LES are compared to RANS, and it is demonstrated that RANS are capable of accurately predicting mean particle impact statistics. However, RANS results display narrower statistical variation than LES, which suggests that particle dispersion is underpredicted in RANS.
		
	\end{abstract}
	

	\section{Introduction}\label{sec:Introduction}
	
	Erosion in gas turbine engines due to surface impact is an important factor contributing to their reduction in performance. Ingestion of small particles such as sand, ash and ice cause harm to the engine, which can eventually lead to engine failure \cite{Tabakoff1996, Tabakoff2006, Clarkson2016}. Erosion is also common in pneumatic conveying systems, such as ducts and pipe elbows, which experience wear due to particle transport \cite{Duarte2016}. The trajectory and size of the particles play an important role in predicting the damage occurring in the engine and pneumatic systems. 
	
	In the past decades, several experimental studies have established that engine degradation occurs due to erosion caused due to particle ingestion. The common methods in conducting these experimental studies is either by impacting particles onto stationary target coupons or by exposing live engines to particles and allowing erosion to occur. The second type of study is less common as it is significant more expensive.
	
	Tabakoff and colleagues have conducted several studies of these studies by injecting particles onto a target plate. The experimental rig developed by Tabakoff utilized a vertical particle acceleration section which was attached to a vertical test section \cite{Tabakoff1974}. Tabakoff et al. have also conducted several investigations into measuring erosion rates on particles impacting specific compressor and turbine materials \cite{Tabakoff1984} with respect to impact angle \cite{Tabakoff1987, Tabakoff1996, Tabakoff1991}, and particle sizes \cite{Tabakoff1975}, as well as target materials \cite{Tabakoff1974, Tabakoff1984, Tabakoff1987, Tabakoff1991, Tabakoff1996}.
	
	Apart from research by Tabakoff and colleagues, several other experimental studies have been conducted, such as the work by Yan et al. \cite{Yan2020}, where they developed their erosion rate prediction model based on tests conducted by impacting particles of various sizes, impact speeds onto a target plate at different angles of attack. Oka and colleagues \cite{Oka2005,Oka2005b} also conducted experimental tests, where they impacted different particles, each with various diameters and impact speeds, onto target plates consisting of different materials. Hufnagel et al. \cite{Hufnagel2018} also conducted various experiments to develop a correlation using dimensional analysis to predict erosion rates, which was compared to the model developed by Oka et al. \cite{Oka2005}.
	
	Furthermore, multiple studies were conducted at Virginia Tech, in which particles of various diameters and speeds were injected into  low speed gas phase flows at various temperatures and impacted onto a target plate at various angles of attack \cite{Boulanger2016, Delimont2014, Delimont2015, Delimont2015a}.
	
	However, many of the studies conducted above implement sand blasting type rigs to inject particles onto the target, and these rigs typically inject high volume fractions of particles through a small diameter nozzle. The simulations conducted in the current study are used to influence experimental tests which have a larger nozzle exit diameter and higher exit speeds compared to those conducted by Delimont et al. \cite{Delimont2014, Delimont2015, Delimont2015a}. 
	
	The long term purpose of this research is to conduct several studies which will help predict the particle dynamics for future experimental research, using the experimental jet rig that is described in the next section, and using that data to predict surface erosion on the compressor section of an engine due to particle impact. A large diameter jet is chosen for these experiments as it enables the experimental research group to conduct various other tests. This preliminary analysis will help in understanding the effect of flow aerodynamics on the particle trajectory and impact behavior in such a test rig.
	
	In this paper we study particles of various sizes being injected onto a flat plate at different velocities, in order to gain an understanding of how particle size affects the impact angle and velocity on a target plate. Another aspect that we study is how particle impacts are affected by varying the target plate angle of attack. As the goal is to drive future experiments, we develop a Computational Fluid Dynamics model of the experimental rig. The experimental rig consists of a constant area duct in which particles are injected in the upstream direction using a particle seeder. The particles travel with the flow through the constant area duct and then exit the rig through a converging nozzle. These particles are impacted onto a target plate which is placed downstream of the converging nozzle. By studying the various particle sizes and target plate angles, we can obtain some statistical information on how the aerodynamic effects induced by the target plate affect the particles. 
	
	\section{Particle Dynamics Characterization}\label{sec:ParticleDynamics}
	
	Particle dynamics in flow can be characterized by their Stokes Number ($St$) which is defined as the ratio of the response time of a particle to the flow characteristic time. The particle response time is given by \cref{eq:part_resp},
	
\begin{equation}
{\tau_p = \frac{\rho_pd^2_p}{18\mu_g}\label{eq:part_resp}}
\end{equation}
	
	For duct flows, the flow characteristic time is typically defined as the ratio of the pipe diameter to the flow velocity, and so the Stokes Number is given by \cref{eq:St}
	
\begin{equation}
St = \frac{\rho_pd^2_pu_g}{18\mu_gd_{\mathrm{rig}}}\label{eq:St}
\end{equation}
	
	There are several corrections for Stokes Numbers, such as studies by Israel and Rosner \cite{Israel1983} where an Effective Stokes Number is used to describe characteristics of particles impacting on to a cylinder, as well as the equations derived by Wessel and Righi \cite{Wessel1988} based on the work by Israel and Rosner \cite{Israel1983}. 
	
	The Stokes Number in this paper is used to characterize the particle diameters and has no influence on the results, and so the classical Stokes Number definition is used to classify the behaviour of the particles.
	
	Particles with $St$ $<<$ 1 act as tracers and follow the flow streamlines whereas particles with $St$ $>>$ 1 detach from the flow and act in a more ballistic manner, while $St$ $\approx$ 1 tend to cluster around vortices \cite{Miranda2020, Hwang2006}.
	
	Since the goal of this paper is to drive experimental tests which uses particle sizes based on real life situations, we choose particles with $St$ ranging from 0.0016 to 62.5, for exit Mach conditions = 0.2, with a focus on particles with $St$ around 7.5 as they are more relevant to particles entering the inlet and compressor of aircraft engines. $St$ for the same particles at $M_{exit} = 0.7$ corresponds to 0.004 to 156.23.

	\section{Numerical And Experimental Setup}\label{sec:Setup}
	
\begin{figure}[H]
		\centering \includegraphics[width=0.9\textwidth]{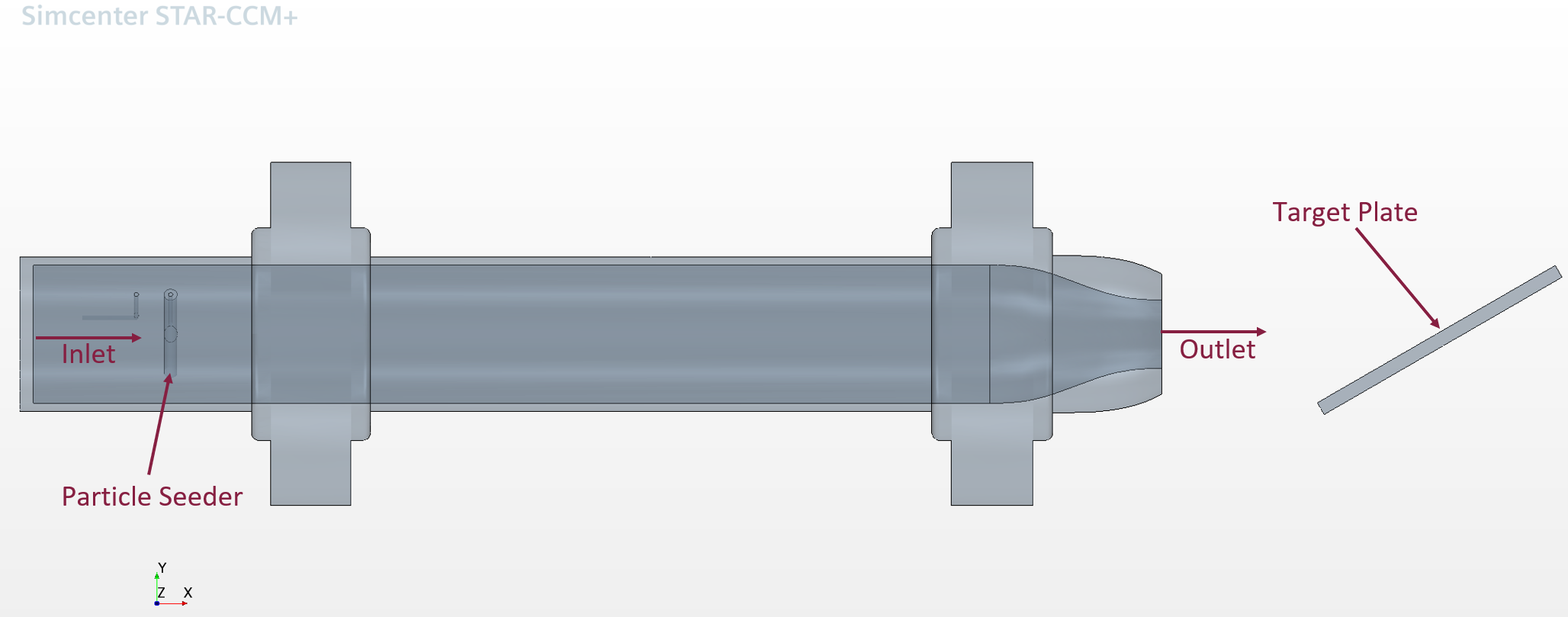}
		\caption{CFD model of the rig and target plate}
		\label{fig:rig} 
	\end{figure}

	In order to reduce computational cost, the CFD model of the rig starts at the location of the final mesh screen, as shown in \cref{fig:rig}.
	
	Another reason for creating the rig model from the final mesh screen is that we can calculate an approximate value of the turbulence intensity ($Tu$) and length scale ($\lambda_x$) based on the mesh screen geometry.
	
	The mesh screens used in the rig have 20 meshes per inch or (20M). Using the information regarding mesh screens provided in \cite{Scheiman1981, DRYDEN1947} as well as turbulence calculations by Roach \cite{Roach1987}, we estimate the turbulence properties listed in Tab.\cref{tab:properties}. 
	
	The experimental rig has a pressure and temperature probe upstream of the particle seeder which would induce turbulence and so the CFD model also has the two probes present as seen in \cref{fig:rig}.
	
	The jet rig has an internal diameter, $d_{\mathrm{rig}} = 0.1016$ $m$, and has a length of 0.8350 $m$ from the inlet to nozzle exit. The converging nozzle has an area ratio of 0.25. 
	
	An external domain needs to be created for the target plate since it is located downstream of the rig nozzle. As the goal of this study is to drive future experiments, it is important to analyze the trajectories of the particles after impact and see if they re-circulate within the room in which the experimental rig is placed. If there is significant re-circulation within the room, additional measures would need to be taken to protect the equipment in the room from repeated erosive impacts from the particles, and so for these CFD simulations the room is considered to be the external domain. The volumetric dimensions of the room are 31x19x20 $ft^3$ (9.45x5.86x6.09 $m^3$).  
	
	\begin{figure}[H]
		\centering \includegraphics[width=0.9\textwidth]{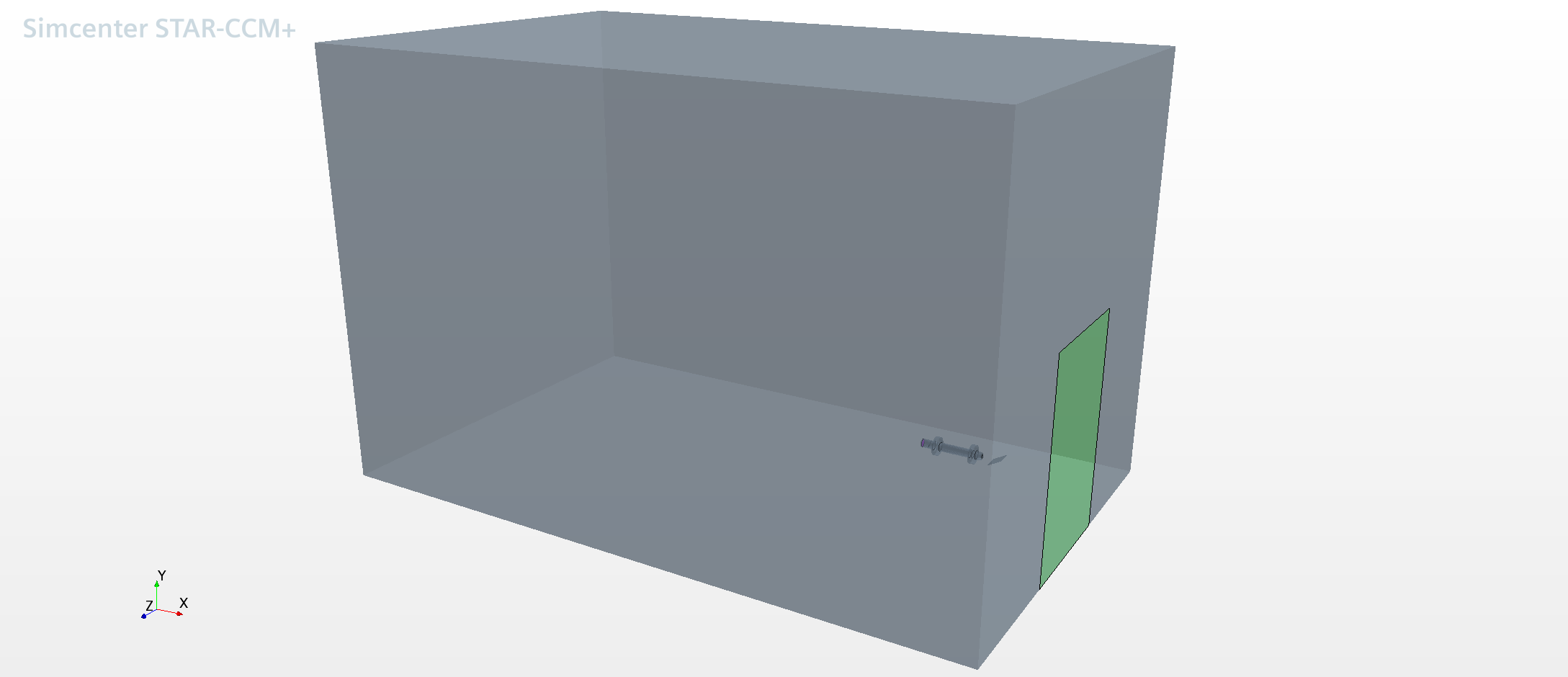}
		\caption{CFD model of the entire simulation domain}
		\label{fig:domain} 
	\end{figure}
	
	LPT-RANS (Lagrangian Particle Tracking - Reynolds-averaged Navier–Stokes) simulations are steady state simulations where the gas phase is solved using RANS models and Lagragian particle tracking is used to solve for the particles in the domain. The flow physics and particle motion are analyzed using LPT-RANS CFD techniques. A 2-equation realizable $k-\varepsilon$ RANS simulation is chosen to model the turbulence physics; the gas phase is two-way coupled to the Lagrangian particle phase. In this study, the two-way coupled model was activated as the volume fraction of particles was approximately 1E-5 \cite{Brandt2022}. The RANS model uses the Two-Layer $k-\varepsilon$ for a Two-Layer All $y^+$ wall treatment. The RANS solver is a pressure based coupled flow solver, where the density is computed from the ideal gas equation.
	
	Studies have shown that there is an agreement between the k-$\varepsilon$ and k-$\omega$ models for time-averaged flow properties. They provide accurate quantitative properties for time averaged flow but are unable to provide information regarding instantaneous flow properties \cite{Vijiapurapu2010}. The k-$\varepsilon$ turbulence model predicts the velocity and friction factor better while the SST k-$\omega$ model gives the most stable and closest estimation of pipe area turbulence intensity compared to other studied turbulence models \cite{Lim2018}, hence the k-$\epsilon$ method is used. The Realizable Two-Layer K-Epsilon model combines the Realizable K-Epsilon model with the two-layer approach. The coefficients in the models are identical, but the model gains the added flexibility of an all- wall treatment \cite{ccmSu}.
	
	The inlet boundary conditions used to set up the StarCCM+ simulation are listed in \cref{tab:properties} below. The door implements an outlet boundary condition where no backflow occurs. And so, the outflow conditions are not prescribed, but they are determined by the flow upstream of the outlet boundary \cite{ccmSu}. 
	
	\begin{table}[H]
		\begin{center}
			\resizebox{0.9\textwidth}{!}{
				\begin{tabular}{c c c c c c c}
					\multicolumn{7}{c}{} \\
					\hline
					\multicolumn{7}{c}{Inlet Flow Conditions}  \\
					\hline
					$M_{exit}$ & $u_{g,exit}$   ($m/s$)& $P_{01}$ (Pa) & $P_1$ (Pa) & $T_{01}$ (K) & $Tu$ ($\%$) & $\lambda_x$ (m) \\
					0.20 & 65 & 9.715E+4 & 9.697E+4 & 285.1 & 15.45 &  1.046E-4 \\
					0.70 & 220 & 1.328E+5 & 1.311E+5 & 285.3 & 15.45 &  1.766E-4 \\
					\hline
					\multicolumn{7}{c}{Particle Seeder Flow Conditions} \\
					\hline
					$\dot{m}_g$ ($kg/s$) & \multicolumn{3}{c}{Particle Mass Flux ($kg/m^2s$)} & \multicolumn{3}{c}{$u_p$ ($m/s$)}\\
					5.913E-3 & \multicolumn{3}{c}{0.6872} & \multicolumn{2}{c}{2.75} \\
					\hline			
			\end{tabular}}
		\end{center}
		\caption{Simulation Flow Properties}
		\label{tab:properties}
	\end{table}
	
	The CFD simulations implement the gravity model to more accurately represent the experimental setup. The drag model implemented is the equation derived by Holzer and Sommerfeld\cite{Holzer2008}. While the shear lift model is the Sommerfeld equation, which expands the shear lift calculation to a wider range of Reynolds numbers than Saffman’s original theory. \cite{ccmSu}. The LPT-RANS simulations also implement a Turbulent Dispersion Model based on the equations by Gosman and Ioannides \cite{Gosman1981, ccmSu}. The normal and tangential components of restitution used in the contact model are both set to their default values, which is 1. Since we are not analyzing any of the current rebound data, there was no emphasis placed on choosing a coefficient of restitution model. The purpose of this study is to understand the aerodynamic behavior of the particles, and how that affects the design of the experimental rig. This is entirely independent of the restitution behavior, so such defaults were deemed appropriate.
	
	\subsection{Grid Study}
	
\begin{figure}[H]
		\centering \includegraphics[width=0.9\textwidth]{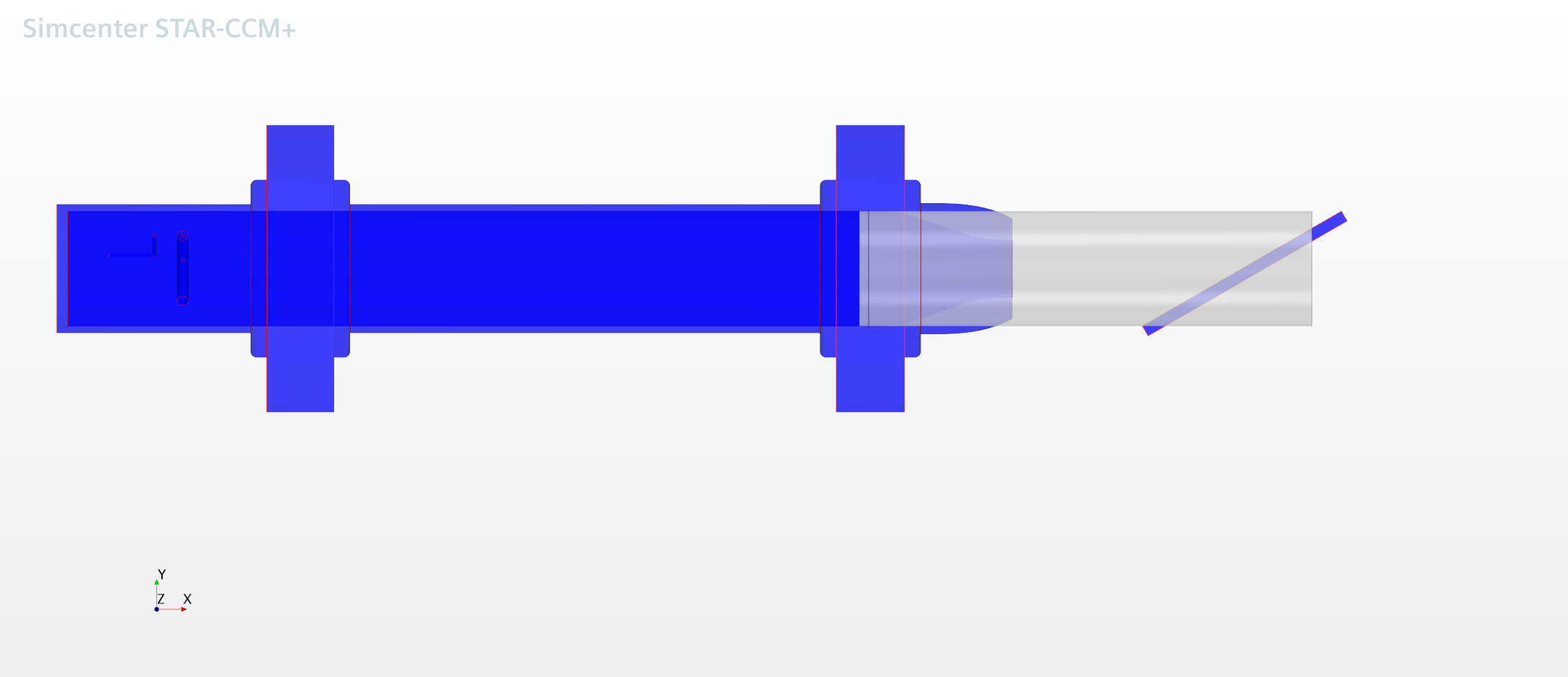}
		\caption{Grid Refinement Region}
		\label{fig:refine}
	\end{figure}

	A shear layer is developed as the flow exiting the nozzle interacts with the stagnant ambient air in the domain. A grid sensitivity study is conducted  to capture the shear layer accurately without the simulations being too computationally expensive. The grid refinement region is a cylindrical region as shown by the grey surface  against blue rig model in \cref{fig:refine}. 
	
	The grid study is conducted using $M_{exit}$ = 0.25 conditions shown in Tab.\cref{tab:properties} The maximum of the magnitude of vorticity is measured within the grid refinement region and plotted as a function of the grid spacing, shown in Tab.\cref{tab:refine}. A Richardson Extrapolation is used to predict the maximum magnitude of vorticity when the grid spacing tends to 0. This shows that the finest grid is capable of capturing the shear layer accurately, and the error between the different grid sizes and the hypothetical grid with 0 spacing are compared in Tab.\cref{tab:refine}. 
	
\begin{figure}[H]
		\centering \includegraphics[width=0.9\textwidth]{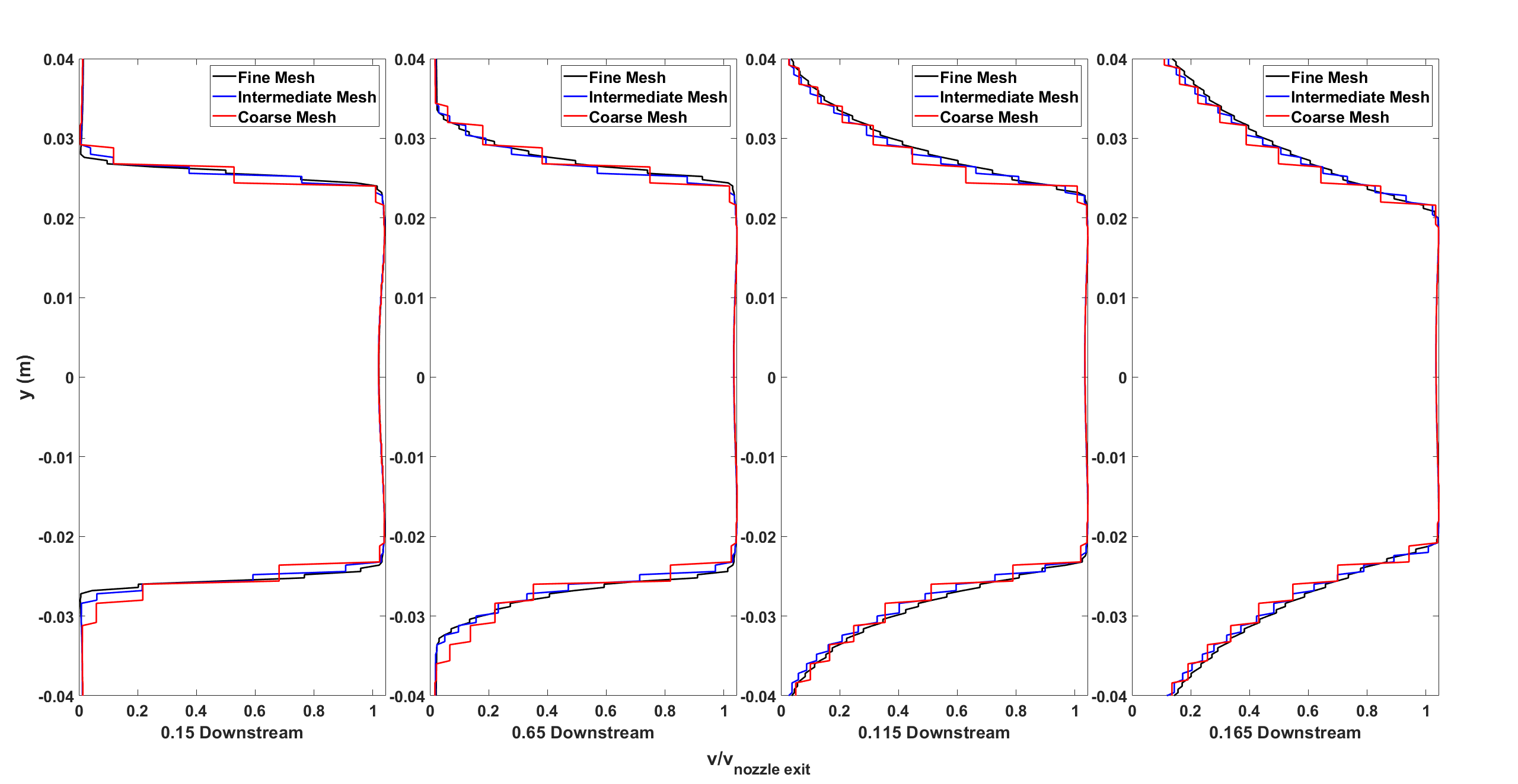}
		\caption{Gas Velocity at locations downstream of the nozzle ($M_{exit} = 0.2$)}
		\label{fig:gas_vel}
	\end{figure}

	This is investigated further by viewing the velocity profiles of the gas velocity at 0.15m, 0.65m, 0.115m, and 0.165m downstream of the nozzle are plotted in \cref{fig:gas_vel}, which shows that all 3 grid configurations are within reasonable accuracy of each other. It is important to note in \cref{fig:gas_vel}, that the raw simulation data  is demonstrated without interpolation. This gives a more accurate understanding of the differences between the meshes, at the cost of introducing a stair-step like artifact into the visualization. The velocity contours capture the shear region where the core jet interacts with the stagnant air. 
	
	The difference between the measured mass flow rates at the rig inlet and exit are measured for the different grid sizes and the error percentage in the mass flow rate between inlet and outlet is shown Tab.\cref{tab:mfr} below. The coarse mesh seems to overpredict the mass flow rate exiting the nozzle while the intermediate and fine grids underpredict it, however all meshes maintain an extremely small error. We performed computational cost analysis on our computer using 70 cores of an Intel Xeon Gold 6242R processor, with 512GB of RAM. Since, the computational time for the fine grid is around 24 hours, while the medium sized grid is capable of achieving convergence in 12 hours, and the relative error between the two grid sizes is small, the medium size grid is used for the simulations conducted in this study.
	
	\begin{table}[H]
		\begin{center}
			\begin{tabular}{ccc}
				\multicolumn{3}{c}{} \\
				\hline
				\thead{Grid \\ Spacing (m)} & \thead{Maximum Magnitude \\ of Vorticity ($s^{-1}$)} & \thead{Relative Error \\ to Finest Mesh ($\%$)} \\
				0 & 4.38E+4 &  0.00 \\
				0.001 & 4.23E+4 &  3.48 \\
				0.002 & 4.05E+4 &  7.51 \\
				0.004 & 3.67E+4 &  16.2 \\
				\hline			
			\end{tabular}
		\end{center}
		\caption{Grid Convergence: Vorticity Error}
		\label{tab:refine}
	\end{table}
		
	\begin{table}[H]
		\begin{center}
			\begin{tabular}{ccc}
				\multicolumn{2}{c}{} \\
				\hline
				\thead{Grid \\ Spacing (m)} &  \thead{Error \\ $\frac{\Delta \dot{m}}{\dot{m}_{in}}$ ($\%$)} \\
				0.001 & -0.0008 \\
				0.002 & -0.0118 \\
				0.004 & 0.0122 \\
				\hline			
			\end{tabular}
		\end{center}
		\caption{Grid Convergence: Mass Flow Error}
		\label{tab:mfr}
	\end{table}
	
	The final grid has approximately 5E+06  mesh cells, and this creates a $y^+$ value of less 5 for $M_{exit}$ = 0.2, and $M_{exit}$ = 0.7 in the constant area duct of the jet rig walls, and $y^+ \approx 10$ in the nozzle section and on the target plate which is suitable for resolving the viscous sublayers using the Two-Layer All $y^+$ wall treatment, while for $M_{exit}$ = 0.7, the $y^+$ value in the nozzle section is around 25, which is towards the upper regime of the buffer region, and the Two-Layer All $y^+$ wall treatment is appropriate to solve for the flow physics in that regime \cite{ccmSu}. \cref{fig:mesh} below displays the refined region along the mid-plane of the jet rig. 
	
	\begin{figure}[H]
		\centering \includegraphics[width=0.9\textwidth]{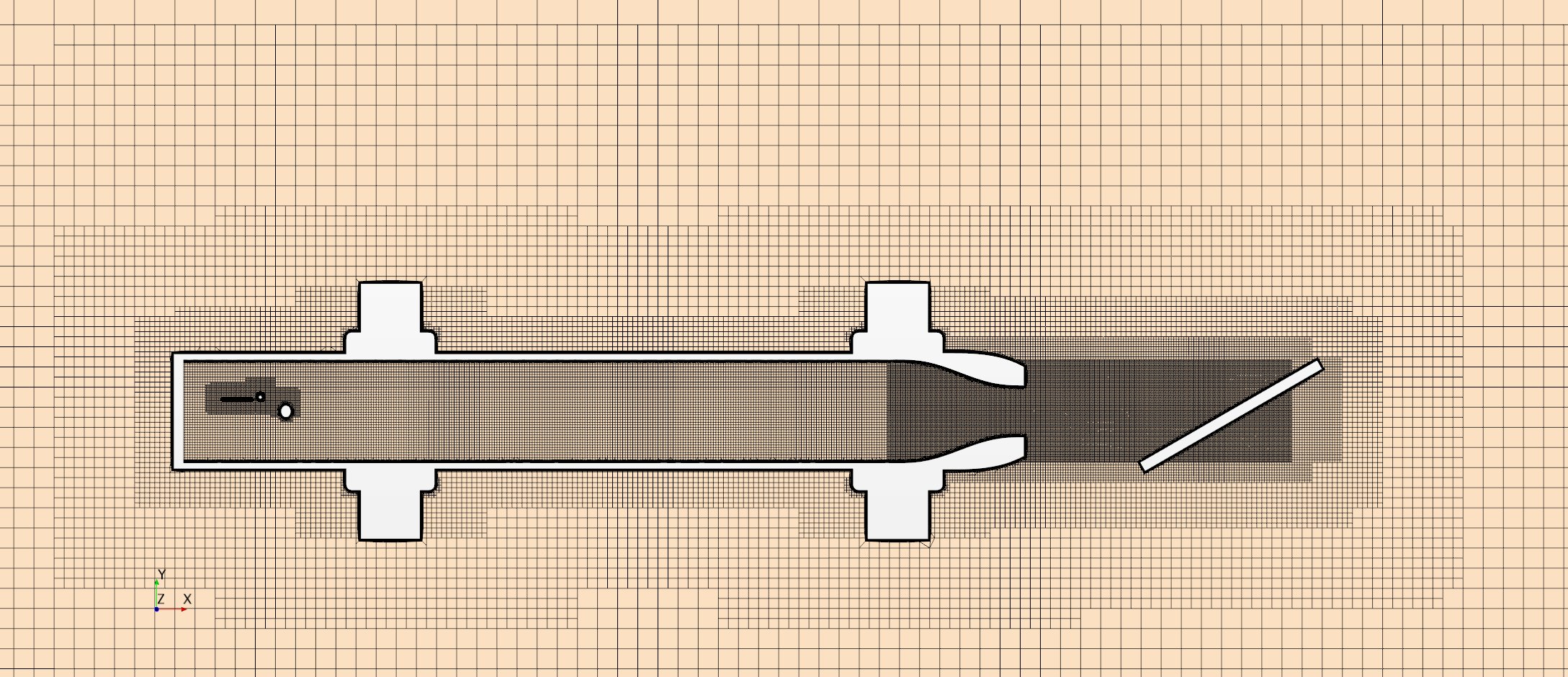}
		\caption{Mesh around the jet rig and target plate setup}
		\label{fig:mesh}
	\end{figure} 
	
	\section{Particle Analysis Simulation Suite}\label{sec:ParticleSuite}
	
	As the goal of this study is to drive future experiments, we setup a suite of simulations to parameterize particle-impact dynamics based on the particle diameter size and orientation of the target plate. The center of the target plate is located at a distance of $2d_{rig}$ downstream of the nozzle. 
	
	Once the RANS simulations achieved fully developed gas phase flow, particles were injected from the particle seeder into the jet rig. The particles are injected upstream into the gas flow so as to minimize particle clustering in the flow. Another reason for injecting the particles upstream, is so their velocity can be independent of the injection mechanism. If they were injected downstream, then the exit velocity would likely depend on the velocity of injection. When injected upstream, their velocities are solely dependent on the gas stream. The particles are tracked from injection to after they rebound and either exit the domain or settle onto the floor due to gravity. Modeling the room and outside domain allows us to track the particles after impact and notice that they do not rebound and impact the target plate after the initial collision.
	
	\subsection{Effects of Particle Diameter on Impact Velocity}
	
	The particle diameters chosen for this study are 1 $\mu m$, 20 $\mu m$, 50 $\mu m$, 70 $\mu m$, 100 $\mu m$, and 200 $\mu m$. This range is chosen to fully span the range of relevant particle sizes seen by the aircraft engine. These particles are injected from the particle seeder and accelerated to impact the target plate which is placed at 30$^\circ$ to the horizontal plane (or the x-axis). A 30$^\circ$ angle is chosen as most compressor blades in gas turbines are placed at this angle.
	
	The boxplot is used for graphically describing a range of data and consists of a box which typically shows the range of data from the first quartile ($25^{th}$ percentile) to the fourth quartile ($75^{th}$ percentile) with a median line at its center. The lines extending vertically from the box depict the maximum and minimum values of the data range.
	
\begin{figure}[H]
		\centering \includegraphics[width=0.9\textwidth]{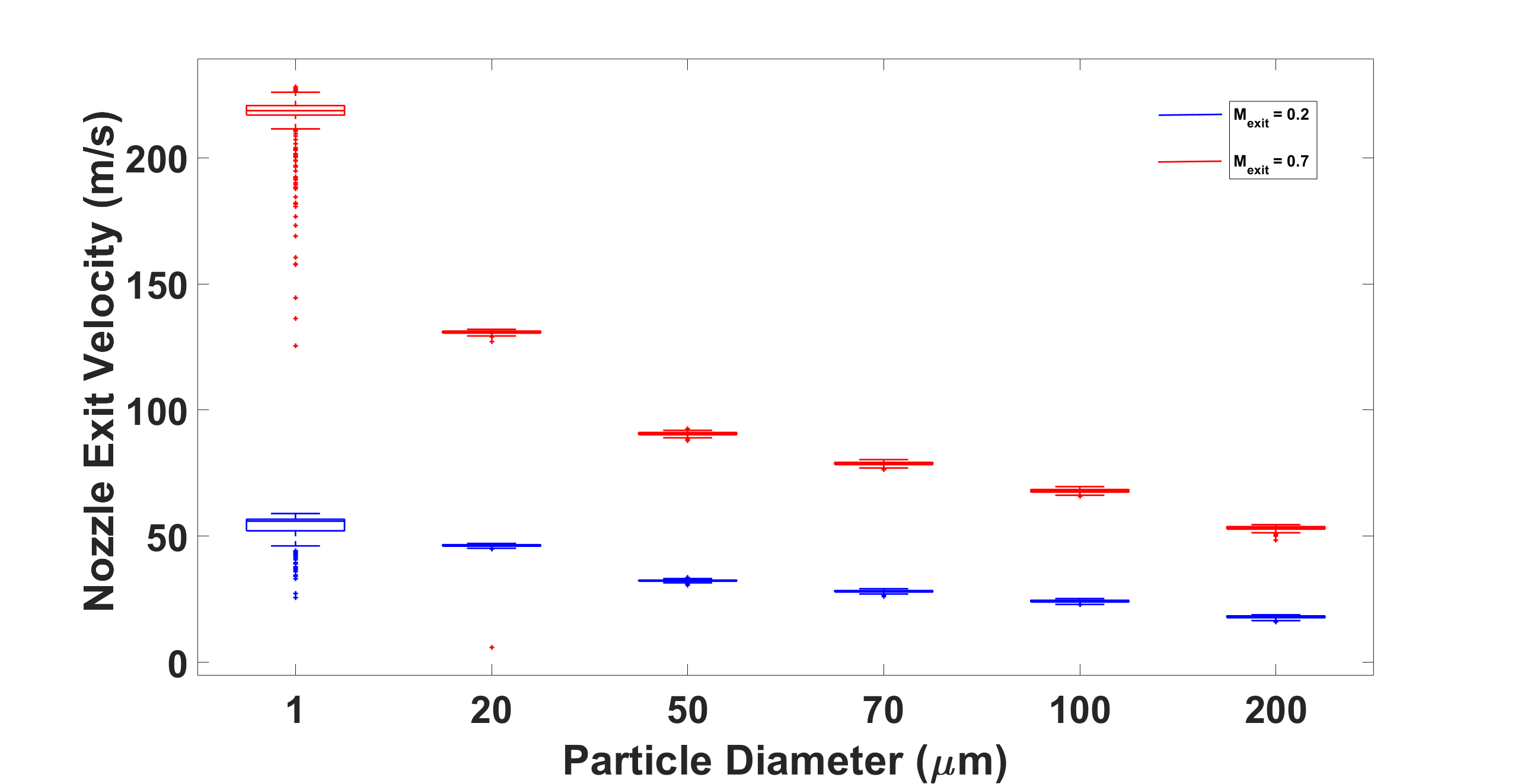}
		\caption{Particle mean exit velocities}
		\label{fig:exit_vel} 
	\end{figure}

	The particle exit velocities are shown in \cref{fig:exit_vel}, which shows a monotonic trend, which is expected, as the smaller particles accelerate faster out of the nozzle while the larger particles have lower nozzle exit velocity. 
	
\begin{figure}[H]
		\centering \includegraphics[width=0.9\textwidth]{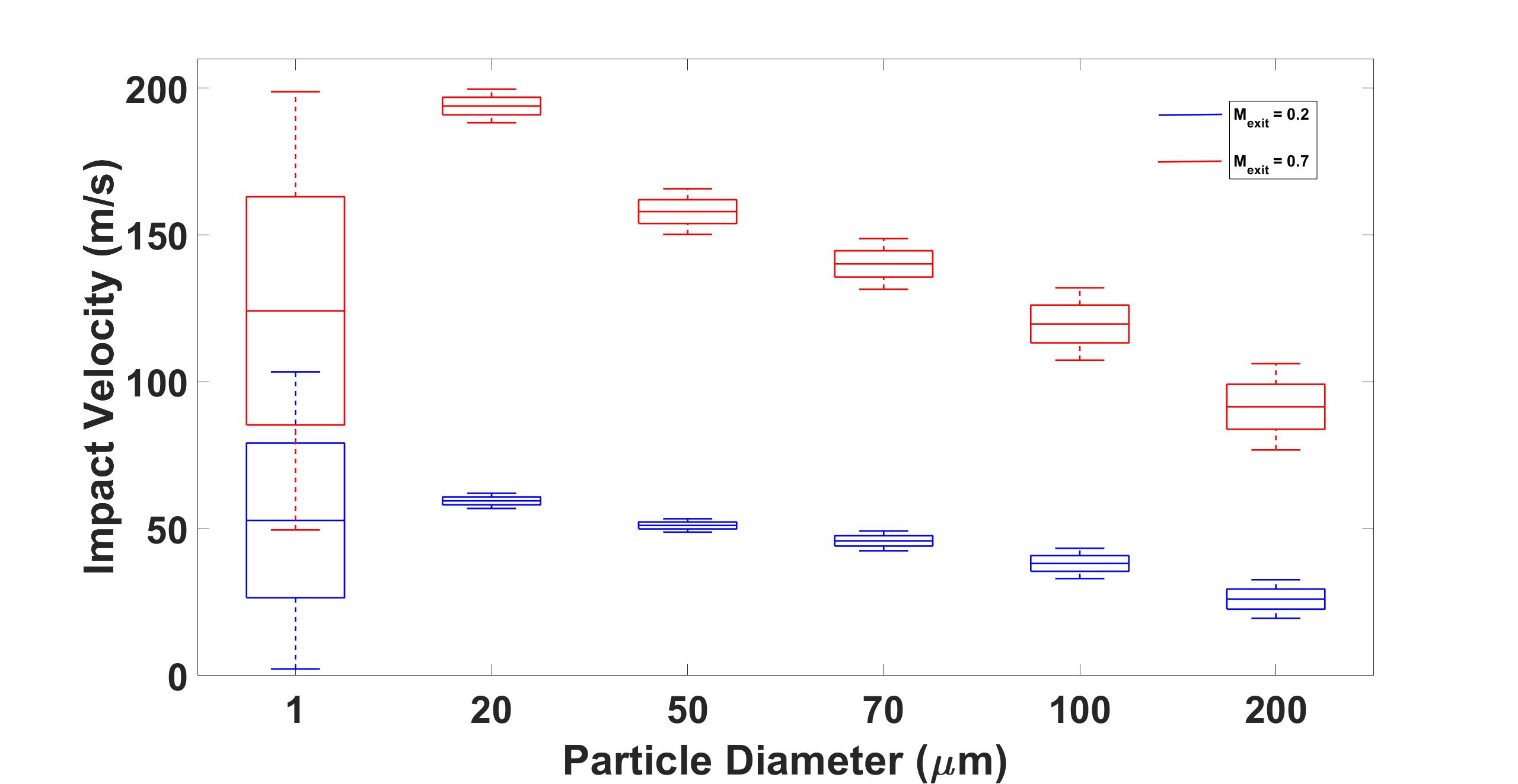}
		\caption{Target Plate Test: Particle impact velocities}
		\label{fig:RANS_vel} 
	\end{figure}

\begin{figure}[H]
		\centering \includegraphics[width=0.9\textwidth]{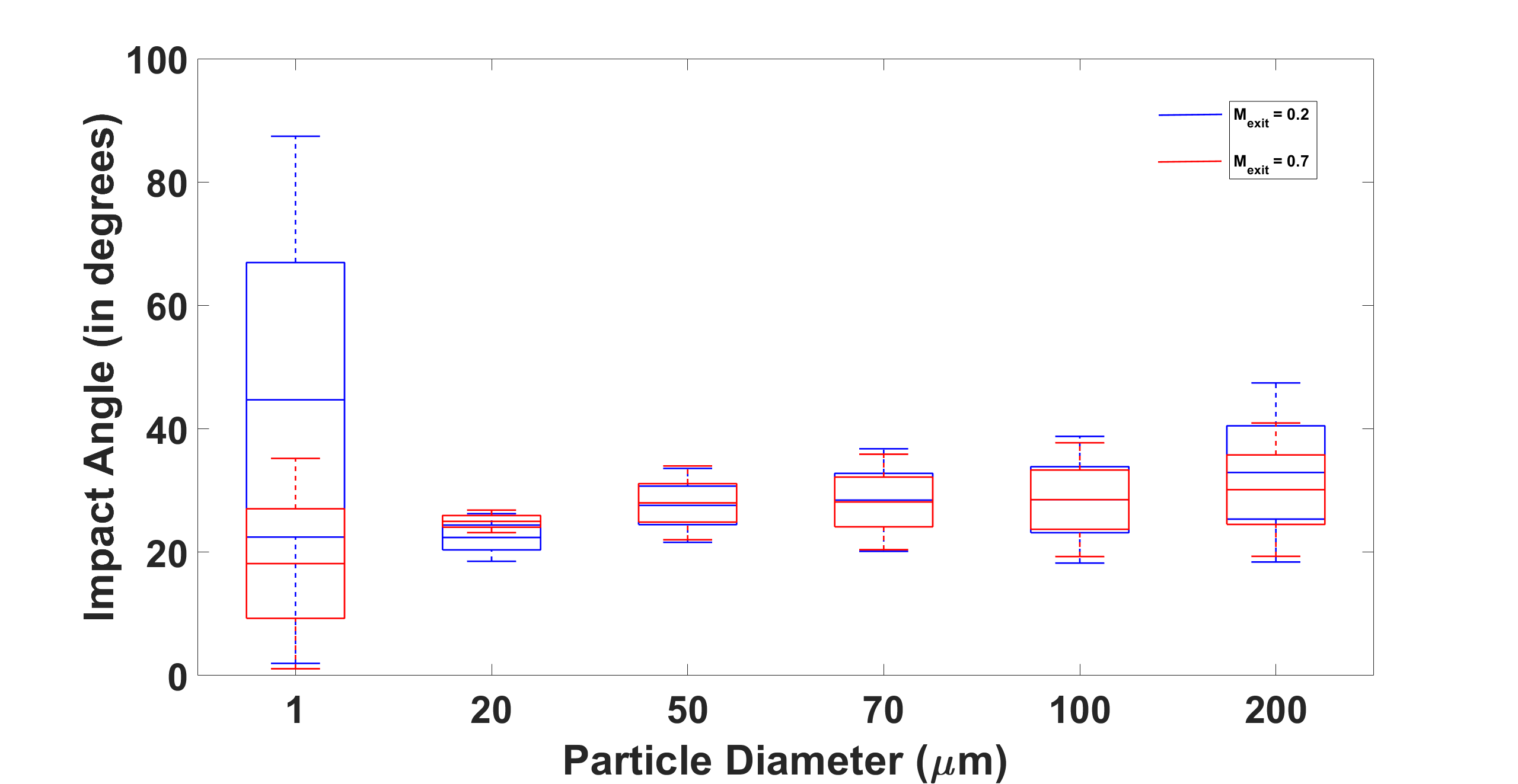}
		\caption{Target Plate Test: Particle incident angles}
		\label{fig:RANS_inc_ang}
	\end{figure}

	A boxplot is used to compare the incidence angles and velocities for the various simulations, which is shown in \cref{fig:RANS_inc_ang} and  \cref{fig:RANS_vel} respectively. It was observed that the mean particle impact velocity follows a non-monotonic trend, which is not intuitive. 
	
	Comparing the mean nozzle exit particle velocities shown in \cref{fig:exit_vel} to the impact velocities in \cref{fig:RANS_vel}, we see an unexpected trend. While the exit velocities are decreasing with St, the impact velocities display a non-monotonic trend. This is more clear by looking at Tab.\cref{tab:exit_vel}
		
	In Tab.\cref{tab:exit_vel} the median of the particle impact velocities from \cref{fig:RANS_vel} is normalized by the median of the particle exit velocities from Tab.\cref{fig:exit_vel}. Table \ref{tab:exit_vel} shows that with the increase in diameter, the ratio $\frac{u_{p,\mathrm{impact}}}{u_{p,\mathrm{exit}}}$ increases upto $d_p = 70 \mu m$, after which $\frac{u_{p,\mathrm{impact}}}{u_{p,\mathrm{exit}}}$ decreases.

	\begin{table}[H]
		
		\begin{center}
			\resizebox{0.9\textwidth}{!}{
				\begin{tabular}{c c c c c c c}
					&  \\ 
					\hline
					\thead{Particle \\ Diameter \\ $\mu m$} & \thead{Exit \\ Velocity \\ $m/s$ ($u_{p,\mathrm{exit}}$)} & \thead{Impact \\ Velocity \\ $m/s$ ($u_{p,\mathrm{impact}}$)} & $\frac{u_{p,\mathrm{impact}}}{u_{p,\mathrm{exit}}}$  & \thead{Exit \\ Velocity \\ $m/s$ ($u_{p,\mathrm{exit}}$)} & \thead{Impact \\ Velocity \\ $m/s$ ($u_{p,\mathrm{impact}}$)} & $\frac{u_{p,\mathrm{impact}}}{u_{p,\mathrm{exit}}}$ \\
					\hline
					1 & 55.95 & 52.82 & 0.94  & 218.7 & 124.1 & 0.56 \\
					20 & 46.38 & 59.46 & 1.28 & 130.8 & 193.8 & 1.48 \\
					50 & 32.37 & 51.10 & 1.57 & 90.58 & 157.9 & 1.74 \\
					70 & 28.10 & 45.84 & 1.63 & 78.73 & 140.1 & 1.78 \\
					100 & 24.26 & 38.17 & 1.57 & 67.85 & 119.7 & 1.76 \\
					200 & 18.09 & 26.04 & 1.44 & 53.33 & 91.5 & 1.71 \\
					\hline
			\end{tabular}}
		\end{center}
		\caption{Non-Dimensionalized Particle Impact Velocities}
		\label{tab:exit_vel}
	\end{table}
	
\begin{figure}[H]
		\includegraphics[width=0.9\textwidth]{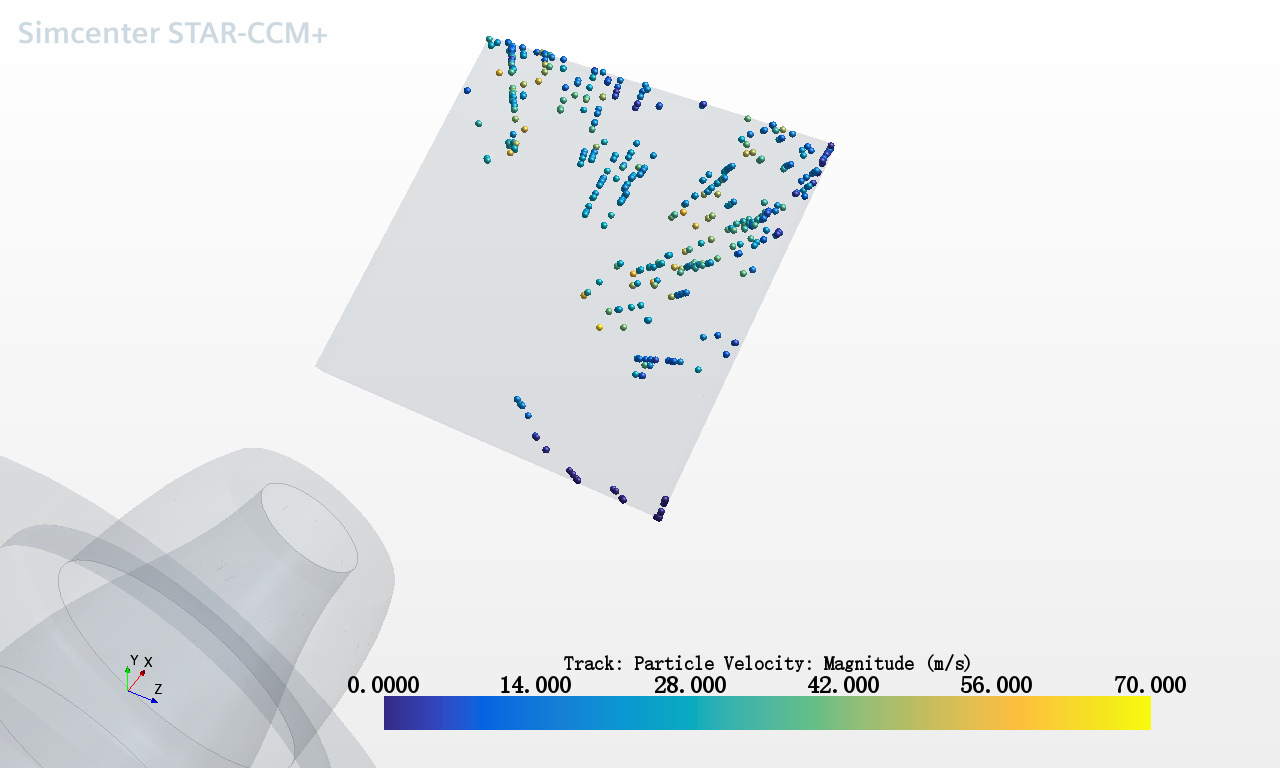}
		\caption{1 $\mu m$ Particle impact ($M_{exit} = 0.2$)}
		\label{fig:1mu} 
	\end{figure}

	We can explain the non-monotonic trend to be due to the physics of particle trajectory. Near the plate, the gas velocity slows down due to the boundary layer development. As the particles exit the nozzle, they accelerate until they approach the plate where they start to slow down (due to boundary layer development and stagnation regions). Since smaller particles respond rapidly to the flow, they also slow down with the gas and disperse over the plate as seen in \cref{fig:1mu}. The larger particles are less susceptible to this sudden change in flow direction and continue to move forward due to inertia. As the larger particles have not reached gas terminal velocity they continue to accelerate until they impact the target plate. It is the interplay of these phenomena that cause the particle impact velocities to be non-monotonic. That is, the impact velocity depends both on exit velocity and the inertia of the particles. This factor must be taken into consideration for the design of future particle impact and erosion rig studies.
	
	As the gas flows over the plate, the 1$\mu m$ particle acts as a tracer and follows the flow streamlines resulting in a large range of incident angles and velocities. The flow speed does not play a significant role in the particle impact angles for the particles with diameters 20 $\mu m$ and larger, however from \cref{fig:RANS_vel} the effect of the flow-plate aerodynamics is stronger for the higher nozzle exit Mach Number. The exit flow speeds play a role in how the 1 $\mu m$ particles impact the plate and so they see a larger variation in both impact angles and velocities.
	
\begin{figure}[H]
		\includegraphics[width=0.9\textwidth]{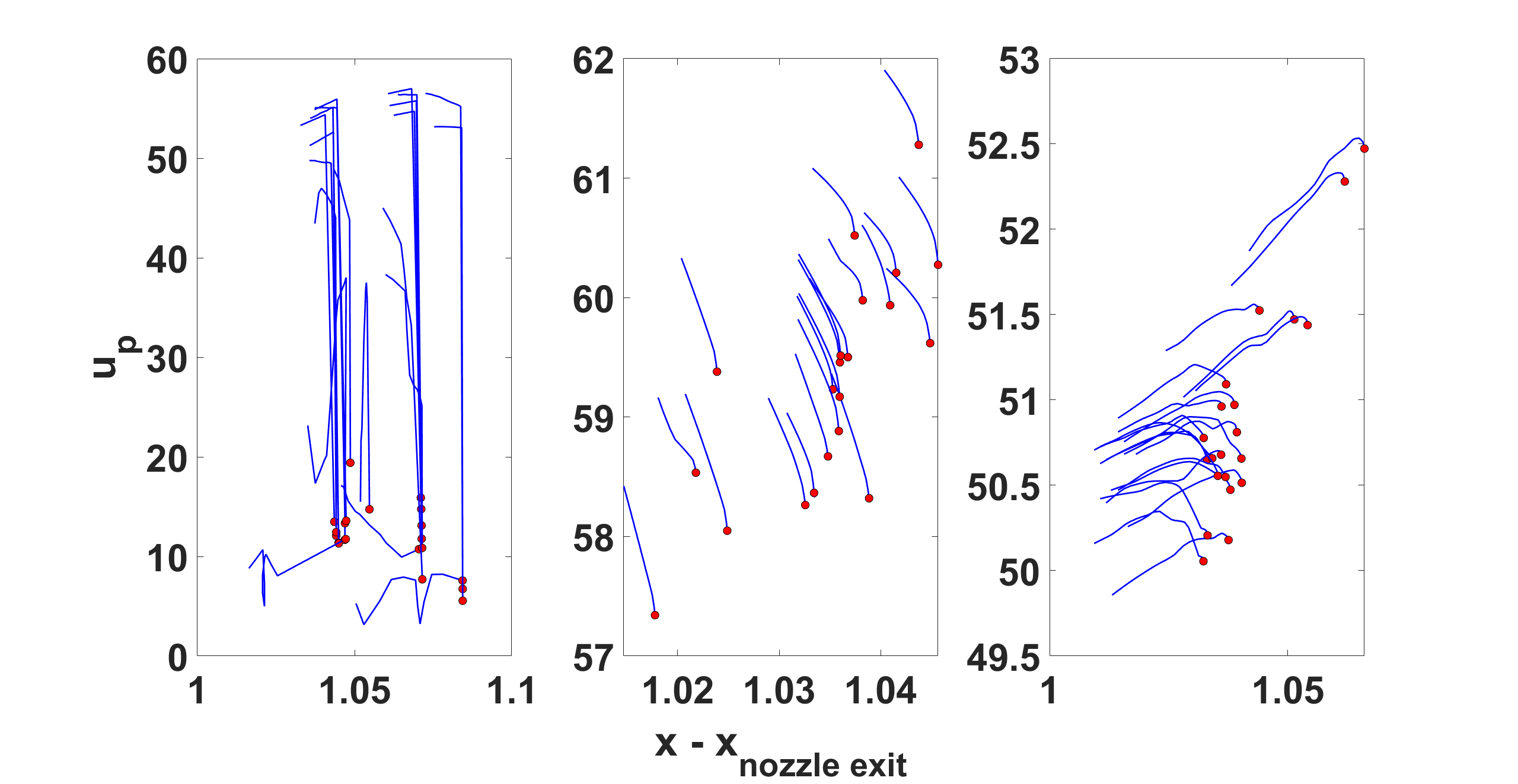}
		\caption{1$\mu m$, 20$\mu m$, and 50$\mu m$ trajectories near the target ($M_{exit} = 0.2$)}
		\label{fig:tracks1} 
	\end{figure}

\begin{figure}[H]
		\includegraphics[width=0.9\textwidth]{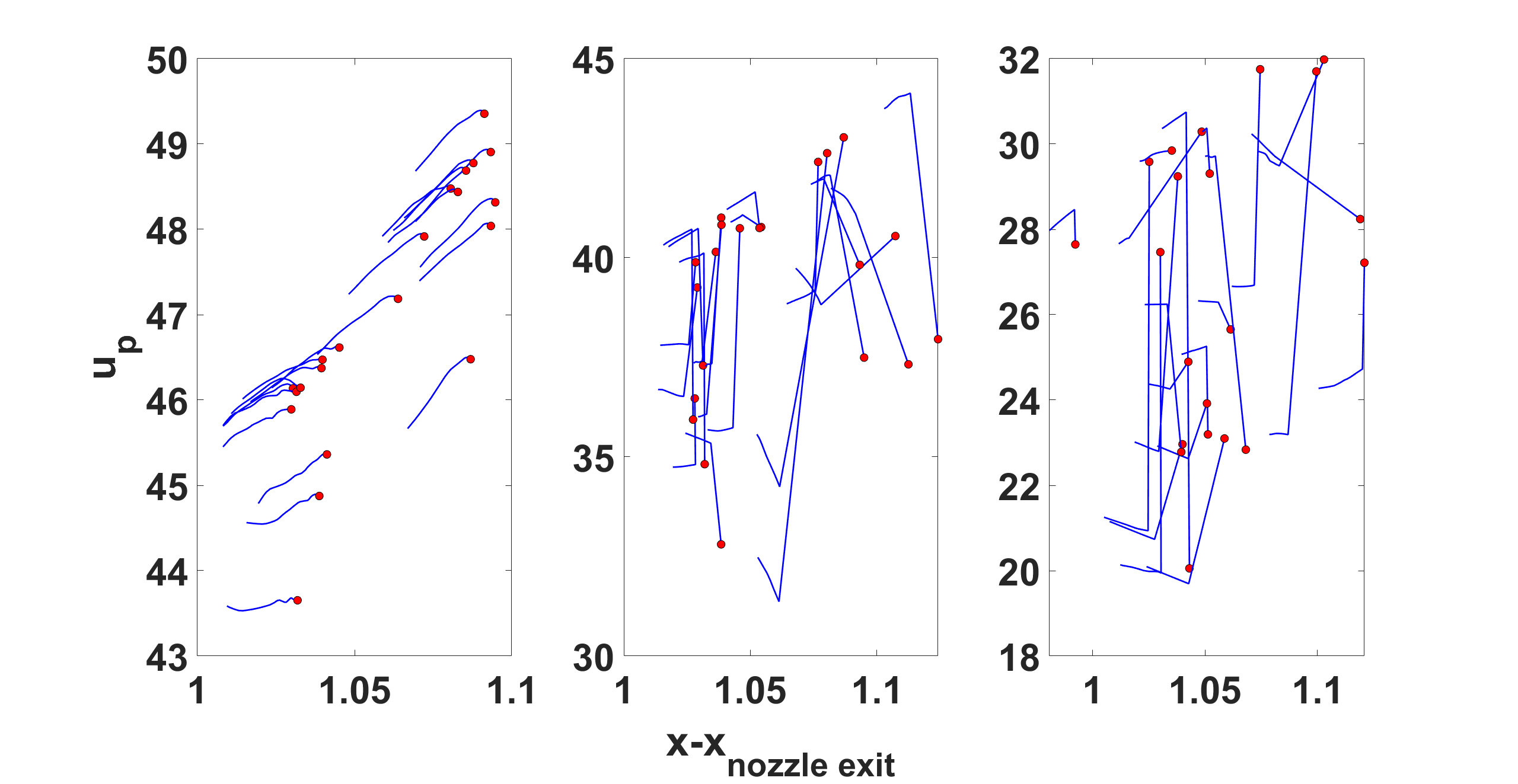}
		\caption{70$\mu m$, 100$\mu m$, and 200$\mu m$ trajectories near the target ($M_{exit} = 0.2$)}
		\label{fig:tracks2} 
	\end{figure}

	\cref{fig:tracks1} to \cref{fig:tracks2} compares the particle velocity magnitudes right before they impact the target plate. It is noted that because particle do not all have the same trajectory or strike at the same location. So several potentially overlapping lines are demonstrated. The images chose to look at 20 particles randomly chosen. As expected, the smallest particles are severely influenced by the flow aerodynamics and boundary layer formation causing them to decelerate rapidly. Larger particles do not follow the flow as much and as they approach the target plate, their speeds reduce due to the region in which the flow interacts with the plate. An example of such a region can be seen later in \cref{fig:stag3}. The influence of this region is seen in \cref{fig:tracks1} to \cref{fig:tracks2} as causing the large variations in velocities seen for each particle size
	
	The influence of the aerodynamic effects of the flow increase when the particles are impacted with a higher nozzle exit Mach Number. The stagnation region that occurs due to the gas flow and target plate interaction causes the particles to slow down much more before impact. Due to the higher inertia of the flow, the 1 $\mu m$ particle trajectories are less erratic and there is a low variation in the range of angles at which they impact the plate
	
	\begin{figure}[H]
		\includegraphics[width=0.9\textwidth]{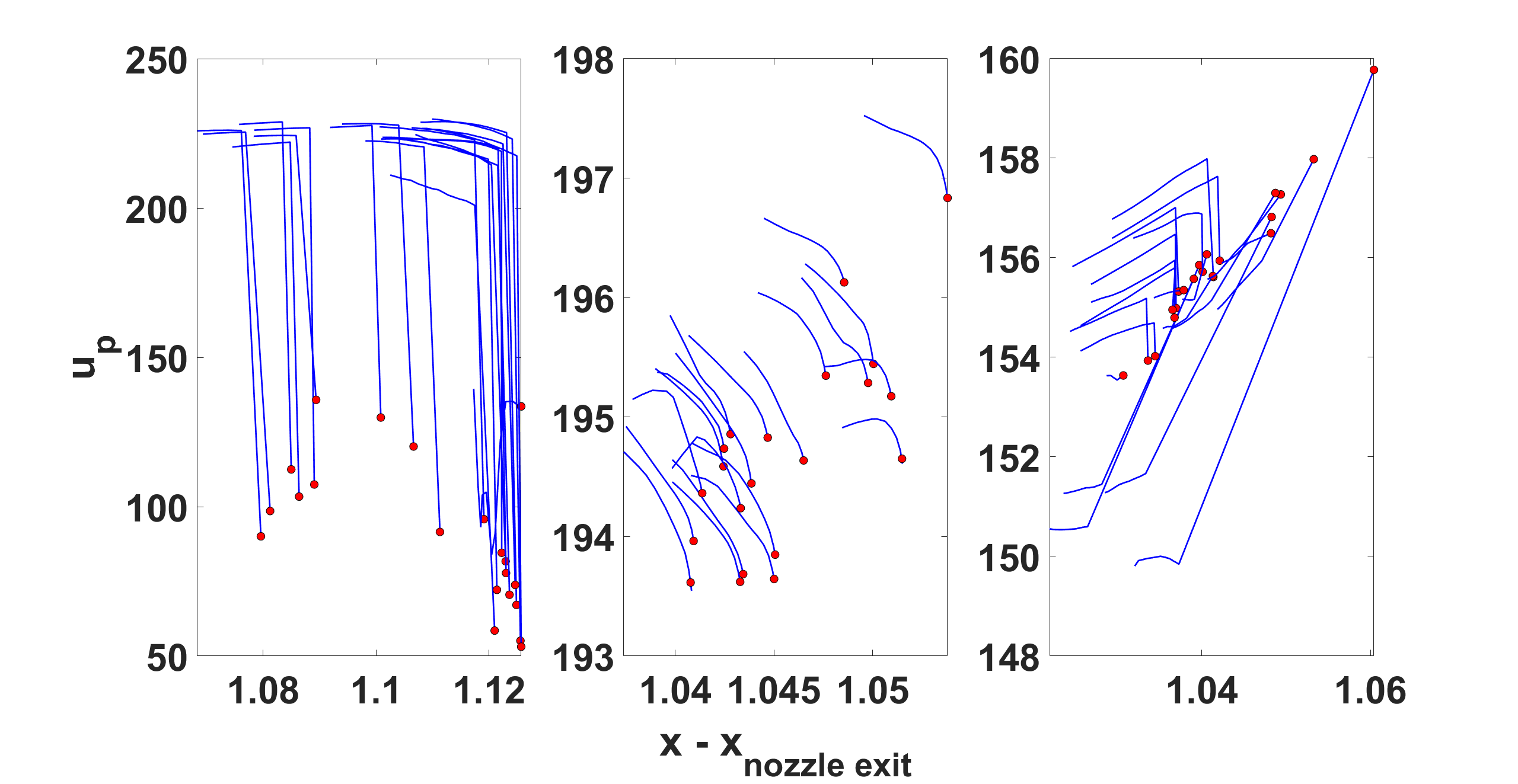}
		\caption{1$\mu m$, 20$\mu m$, and 50$\mu m$ trajectories near the target ($M_{exit} = 0.7$)}
		\label{fig:tracks3} 
	\end{figure}
	
	\begin{figure}[H]
		\includegraphics[width=0.9\textwidth]{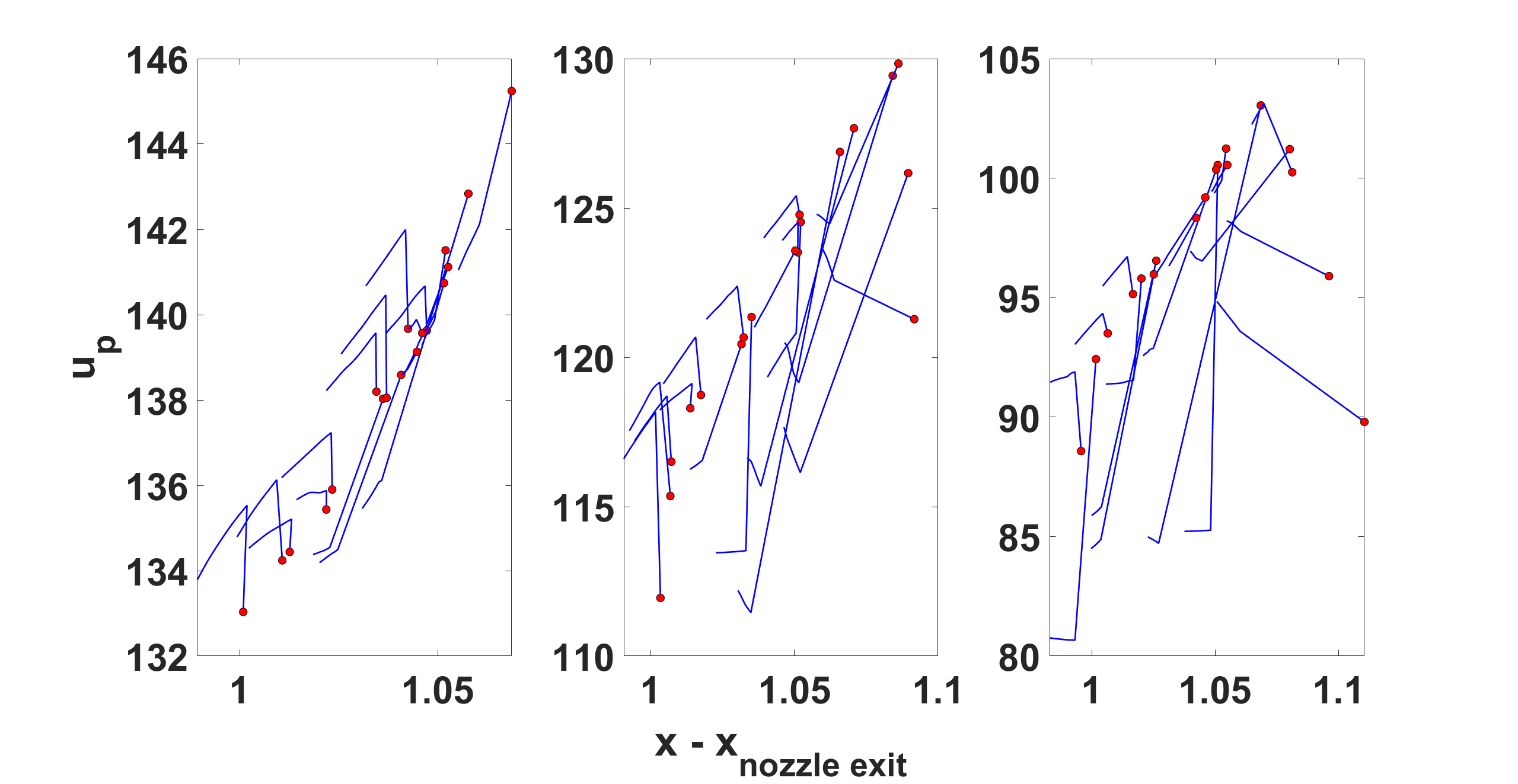}
		\caption{70$\mu m$, 100$\mu m$, and 200$\mu m$ trajectories near the target ($M_{exit} = 0.7$)}
		\label{fig:tracks4} 
	\end{figure}
	
	The flow around the stagnation region acts as a ``slingshot'' for the largest particles. From \cref{fig:tracks1} to \cref{fig:tracks4}, we see that some of the larger particle trajectories show an increase in velocity before contact with the target plate. This is consistent with the thesis that large particles continue accelerating even after the duct exit. However, very near to the target plate, they demonstrate strong deceleration caused by the interaction with the boundary layer
	
\begin{figure}[H]
		\includegraphics[width=0.9\textwidth]{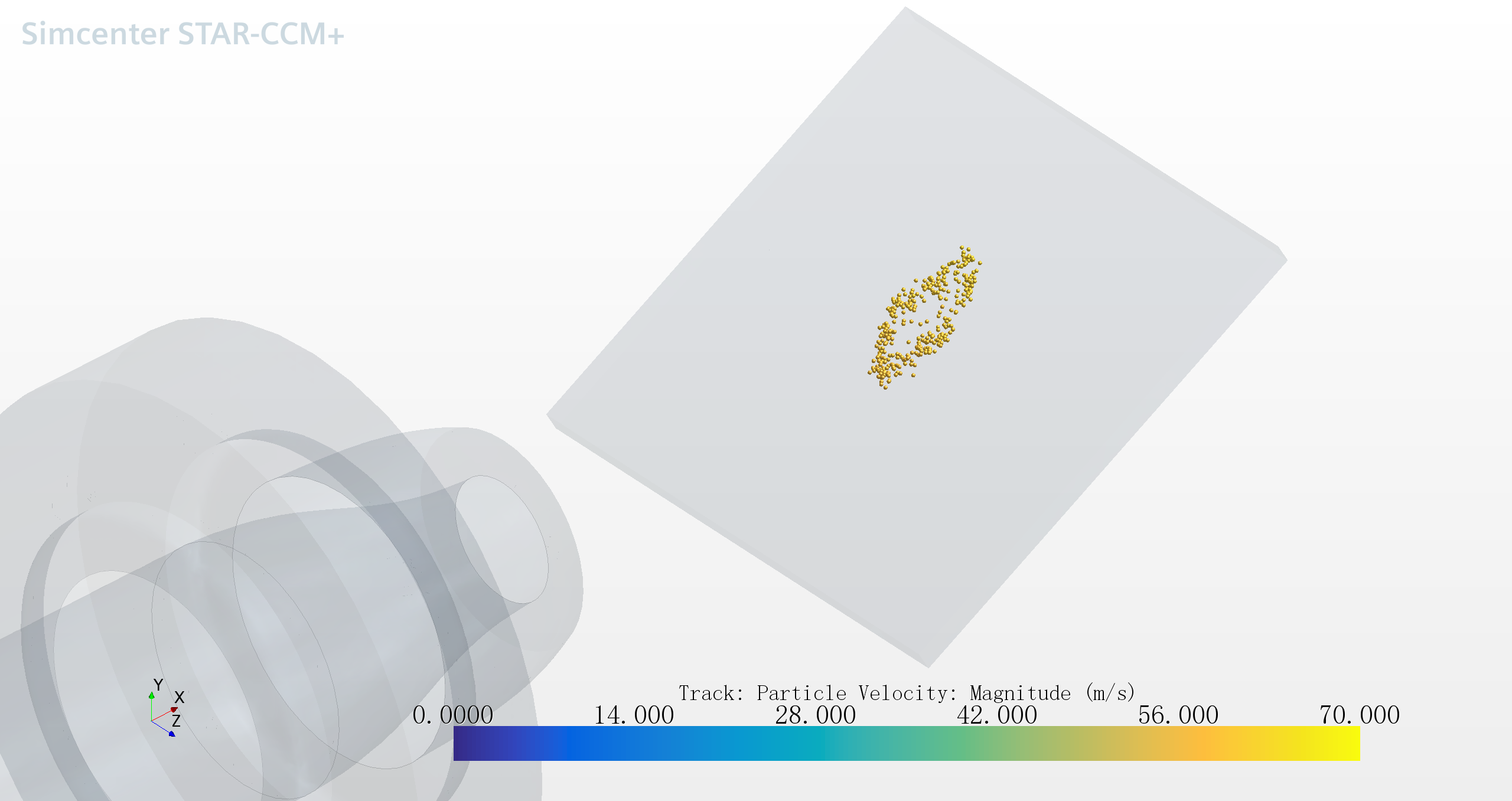}
		\caption{20 $\mu m$ Particle impact ($M_{exit} = 0.2$)}
		\label{fig:20mu} 
	\end{figure}

\begin{figure}[H]
		\includegraphics[width=0.9\textwidth]{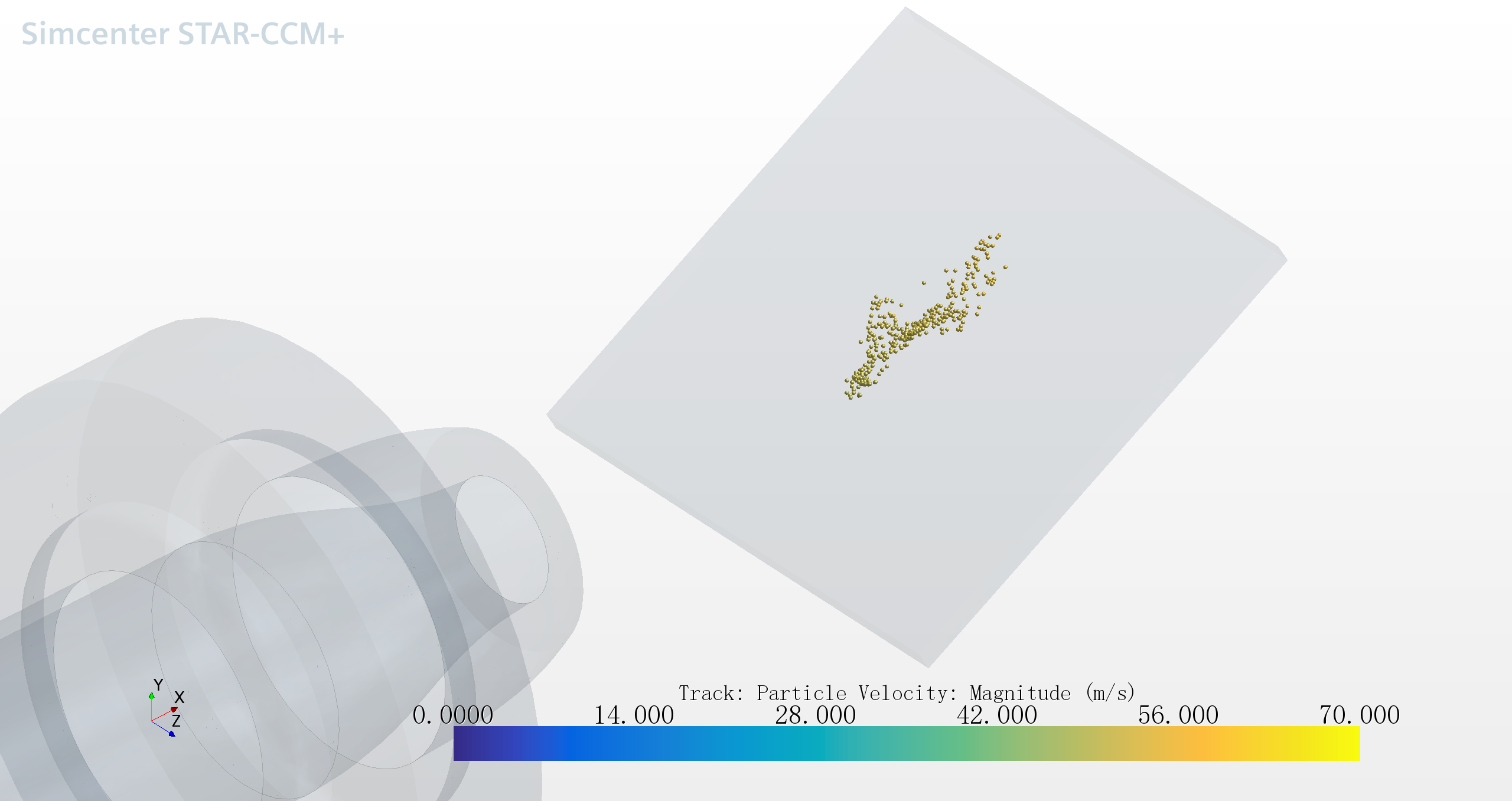}
		\caption{50 $\mu m$ Particle impact ($M_{exit} = 0.2$)}
		\label{fig:50mu} 
	\end{figure}

\begin{figure}[H]
		\includegraphics[width=0.9\textwidth]{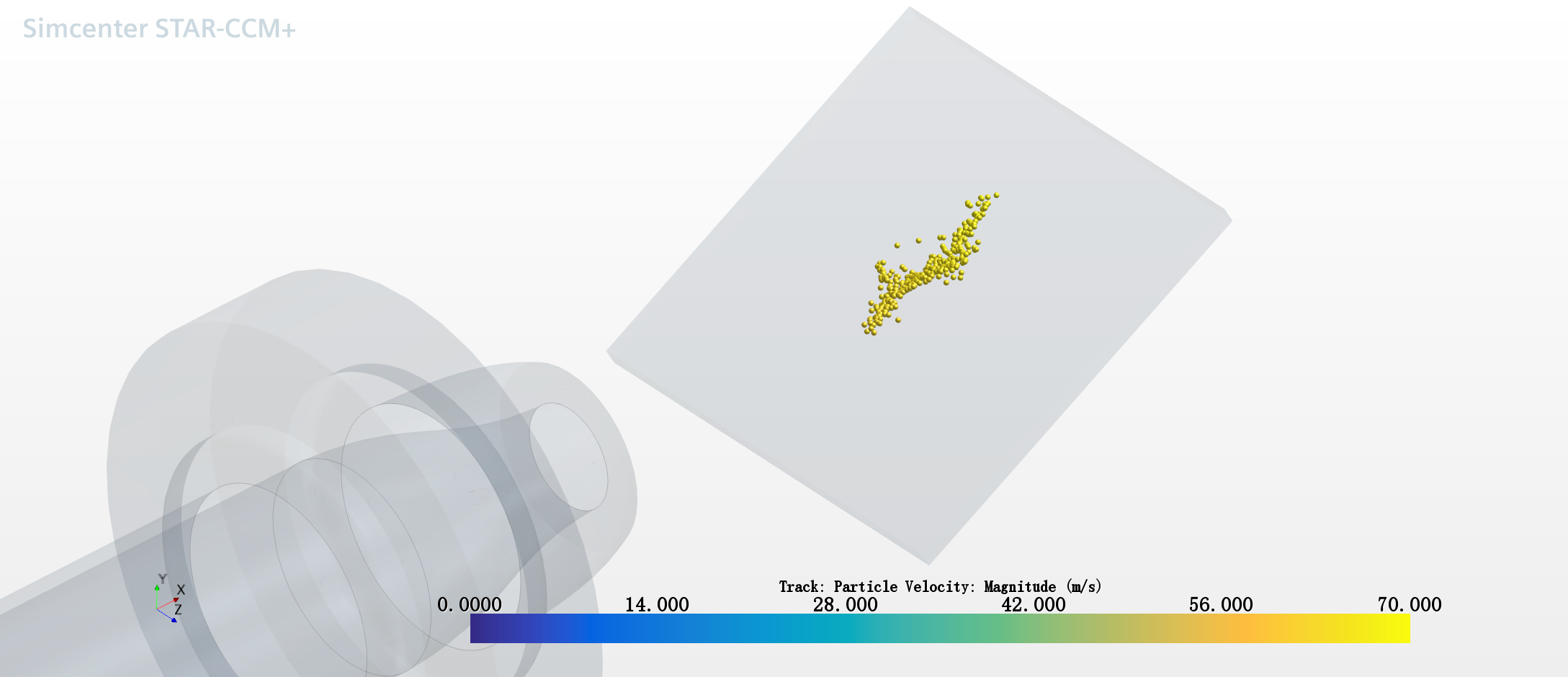}
		\caption{70 $\mu m$ Particle impact ($M_{exit} = 0.2$)}
		\label{fig:70mu} 
	\end{figure}

\begin{figure}[H]
		\includegraphics[width=0.9\textwidth]{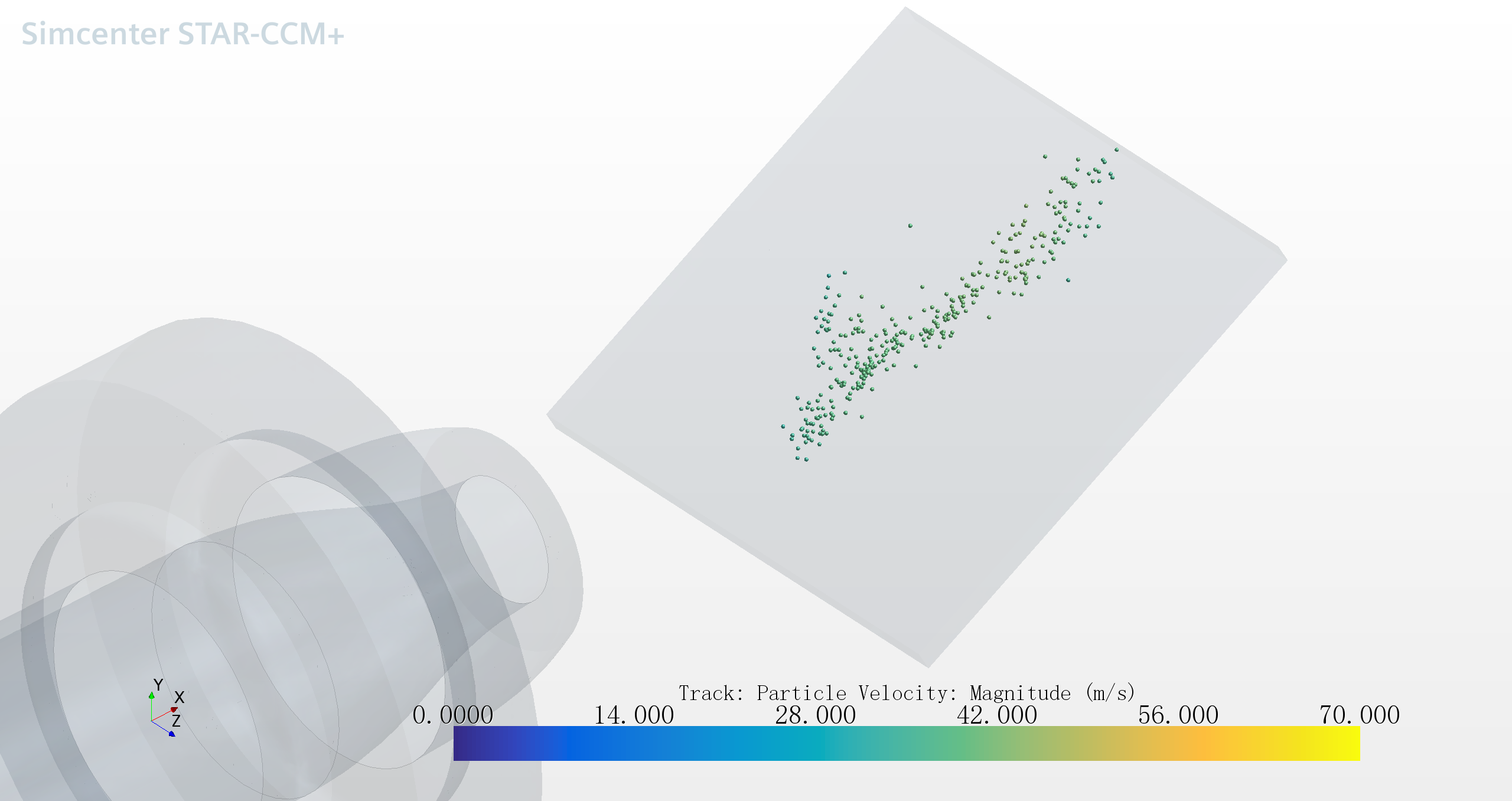}
		\caption{100 $\mu m$ Particle impact ($M_{exit} = 0.2$)}
		\label{fig:100mu} 
	\end{figure}

\begin{figure}[H]
		\includegraphics[width=0.9\textwidth]{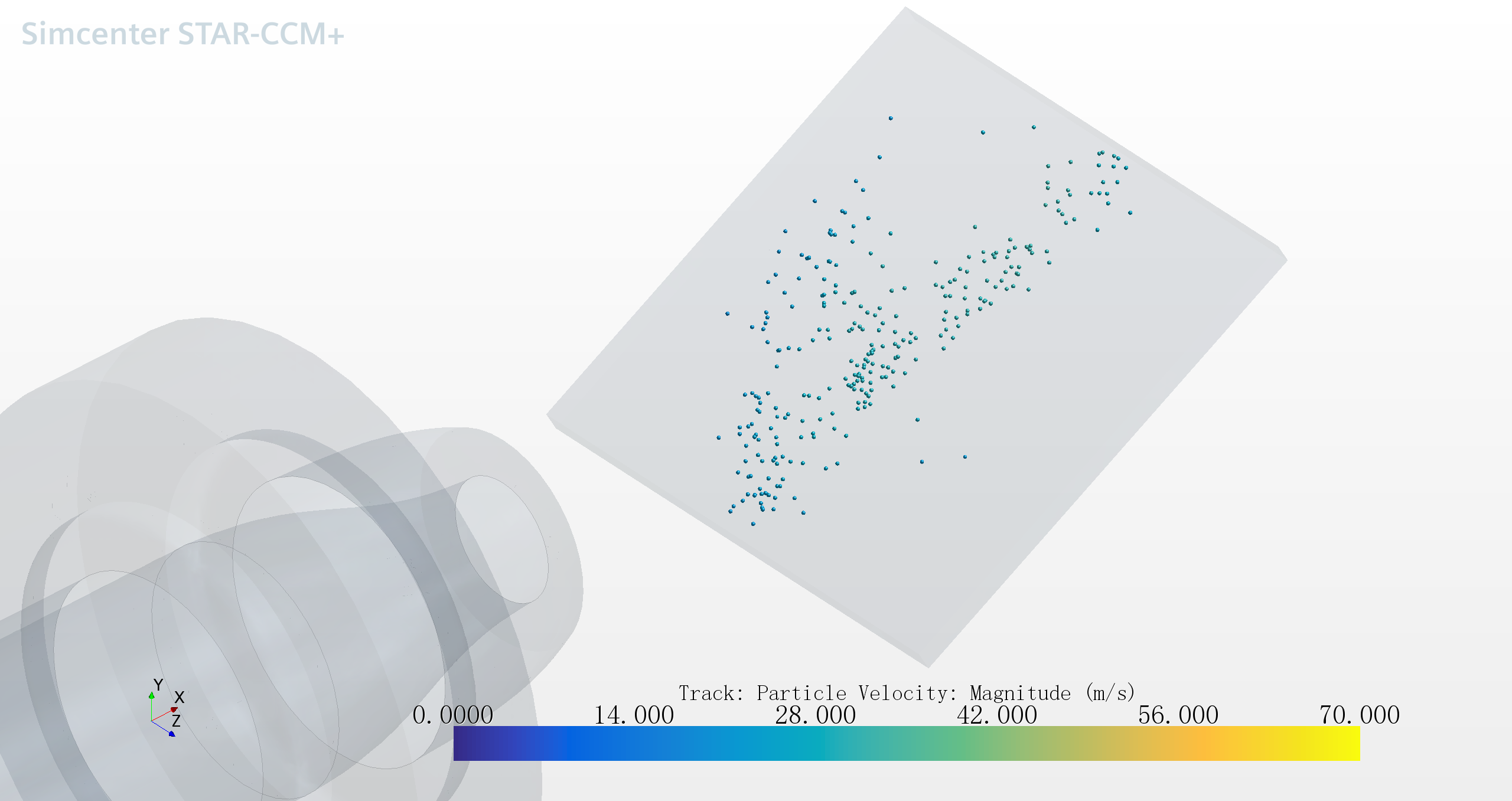}
		\caption{200 $\mu m$ Particle impact ($M_{exit} = 0.2$)}
		\label{fig:200mu} 
	\end{figure}

	 \cref{fig:1mu} to \cref{fig:200mu} compare the particle impact locations based on sizes, which helps understand the large range of incidence angles and velocities for particles of 1 $\mu m$ diameter. The difference between the impact locations with $M_{exit} = 0.2$ and $M_{exit} = 0.7$ is negligible and so \cref{fig:1mu} to \cref{fig:200mu} are the results from the $M_{exit} = 0.2$ study. As the gas flows over the plate and interacts with the boundary layer, the smallest particle ($St_{\mathrm{eff}} = 0.0013$) acts as a tracer and follows the flow streamlines resulting in a large range of incident angles and velocities. The larger particles have more inertia and thus impact the plate in a less diffuse manner. Interestingly, the largest particles also experience some degree of diffusion. This phenomenon is related to randomness in the distribution of particle velocities before they exit the main duct. As particles rebound from the duct walls, their initial velocity at duct exit is not uniform, and this effect is felt more intensely for larger, more ballistic particles.
	
	\subsection{Effects of the Angle of the Target Plate}
	
	The target plate is rotated from 10$^\circ$ to 60$^\circ$ to analyze the effects of particle clustering and impact velocity. Particle diameters ranging from 10 $\mu m$ to 150 $\mu m$, are the most relevant for this study and so particles with 70 $\mu m$  diameter are chosen. These are also the target size of particles used in the preliminary experiments at Virginia Tech.
	
\begin{figure}[H]
		\centering \includegraphics[width=0.9\textwidth]{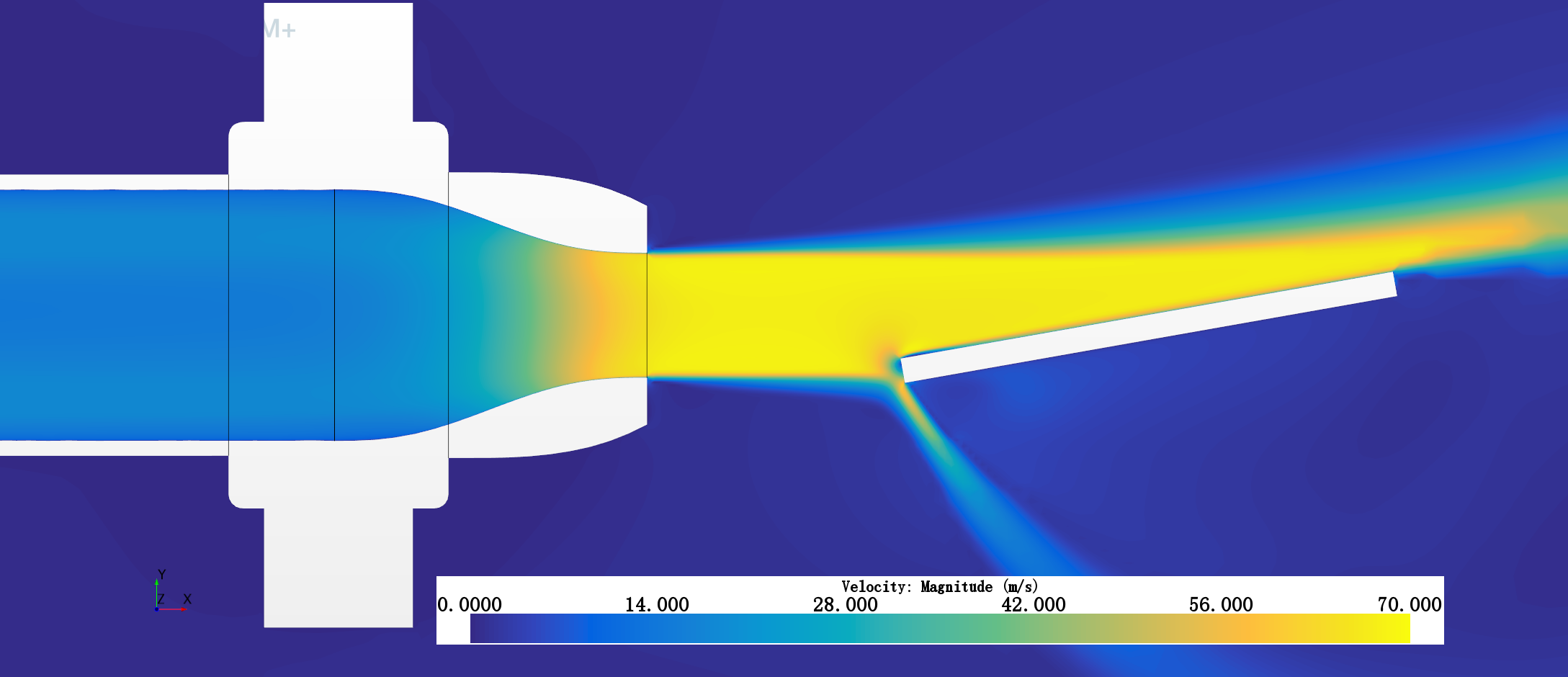}
		\caption{Velocity contour showing a stagnation region ($10^\circ$ angle of attack, $M_{exit} = 0.2$)}
		\label{fig:stag1} 
	\end{figure}

\begin{figure}[H]
		\centering \includegraphics[width=0.9\textwidth]{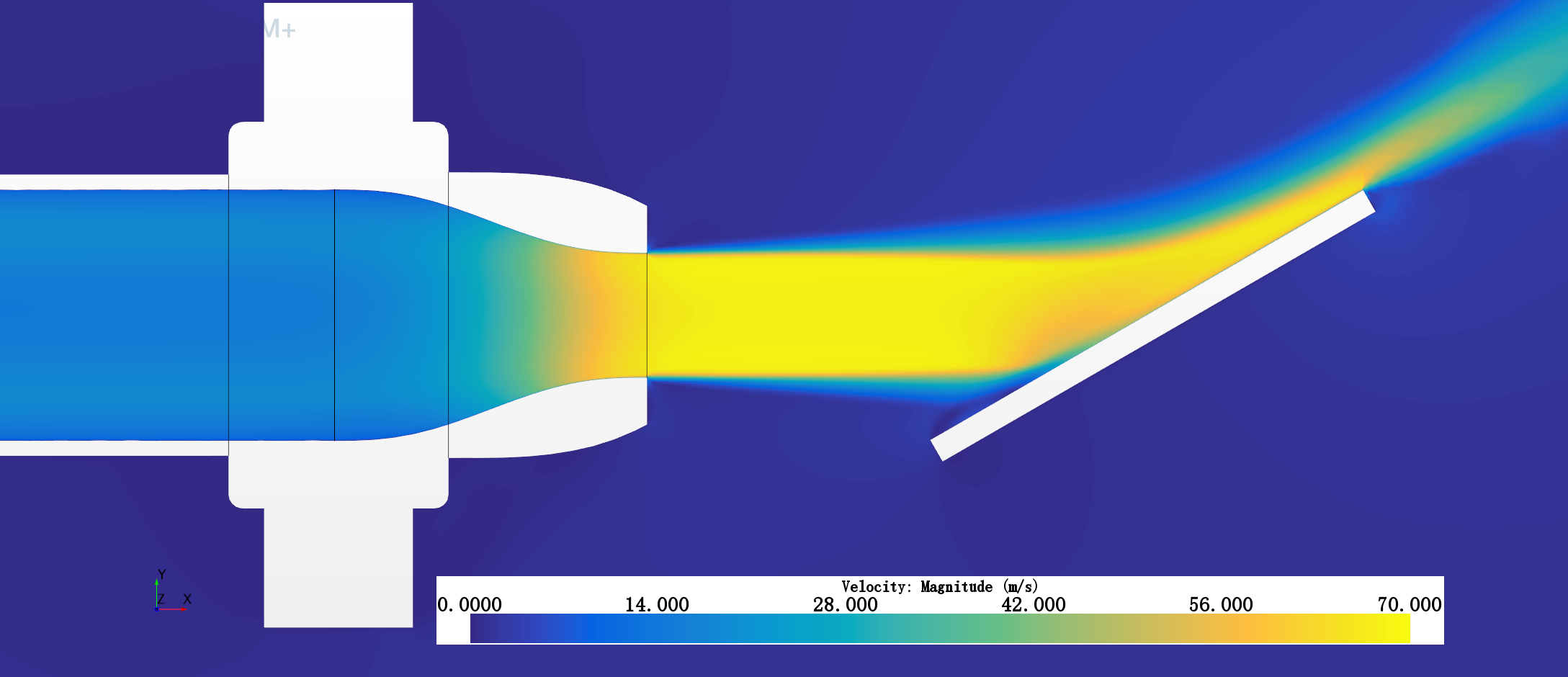}
		\caption{Velocity contour showing a stagnation region ($30^\circ$ angle of attack, $M_{exit} = 0.2$)}
		\label{fig:stag2} 
	\end{figure}

\begin{figure}[H]
		\centering \includegraphics[width=0.9\textwidth]{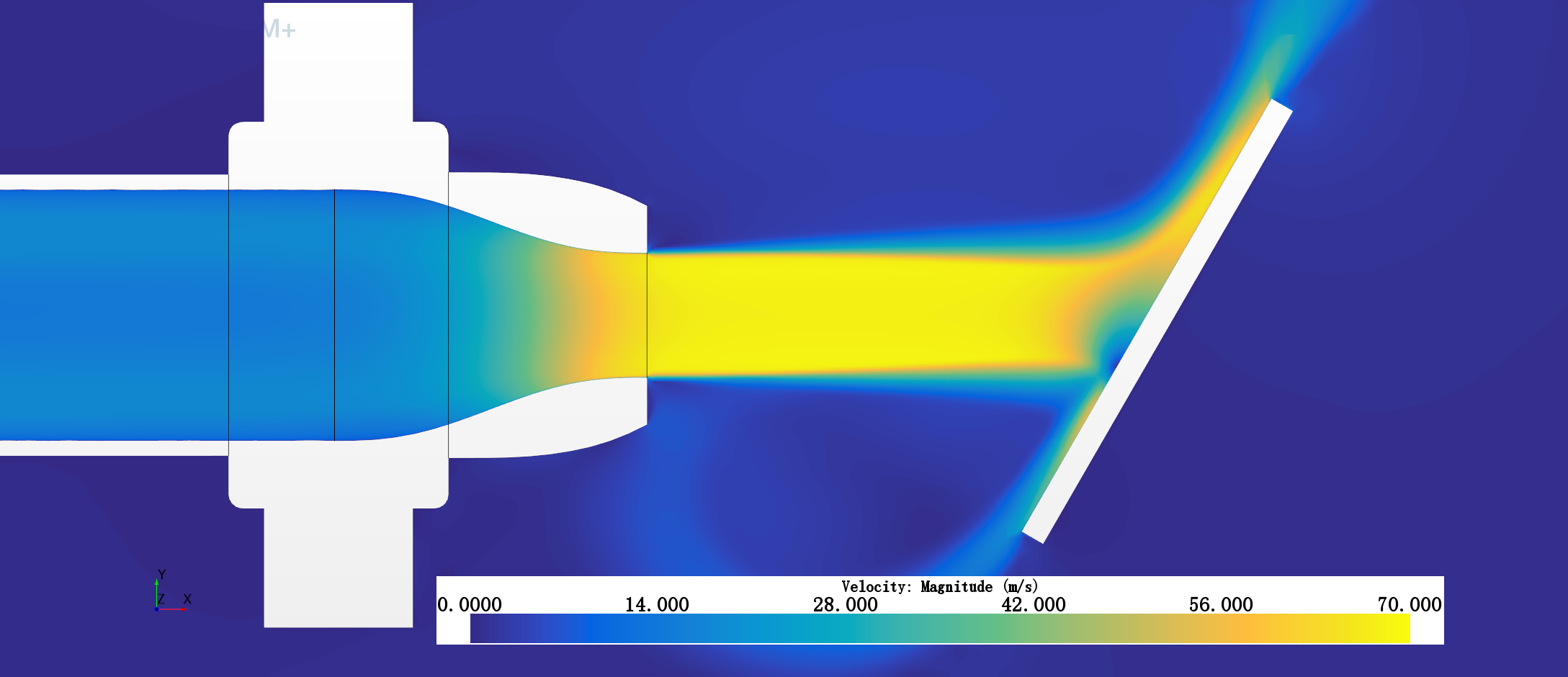}
		\caption{Velocity contour showing a stagnation region ($60^\circ$ angle of attack, $M_{exit} = 0.2$)}
		\label{fig:stag3} 
	\end{figure}

\begin{figure}[H]
		\centering \includegraphics[width=0.9\textwidth]{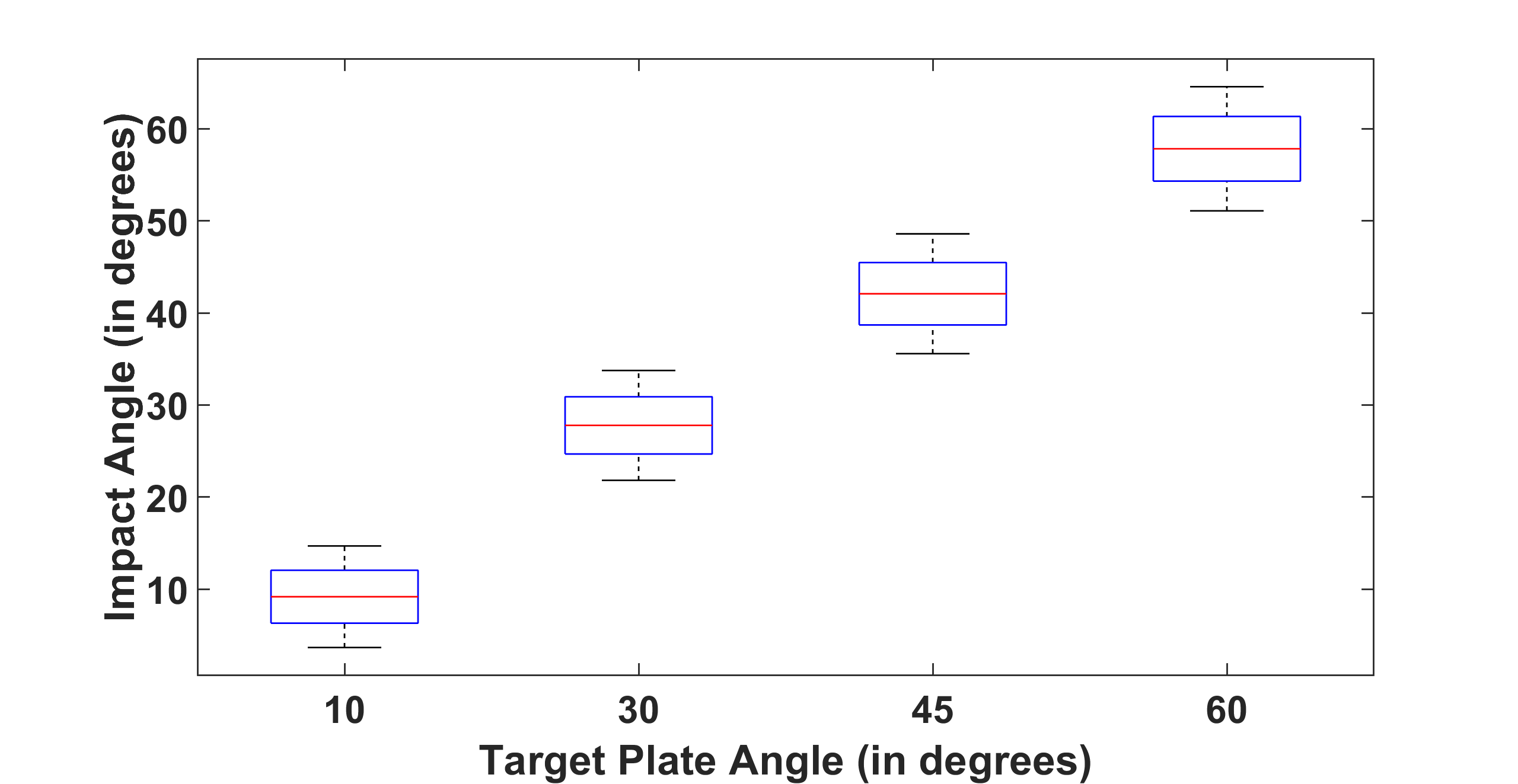}
		\caption{Target Plate Angle Test: Particle incident angles  (70 $\mu m$ diameter particles, $M_{exit} = 0.2$)}
		\label{fig:ang_inc_ang} 
	\end{figure}

\begin{figure}[H]
		\centering \includegraphics[width=0.9\textwidth]{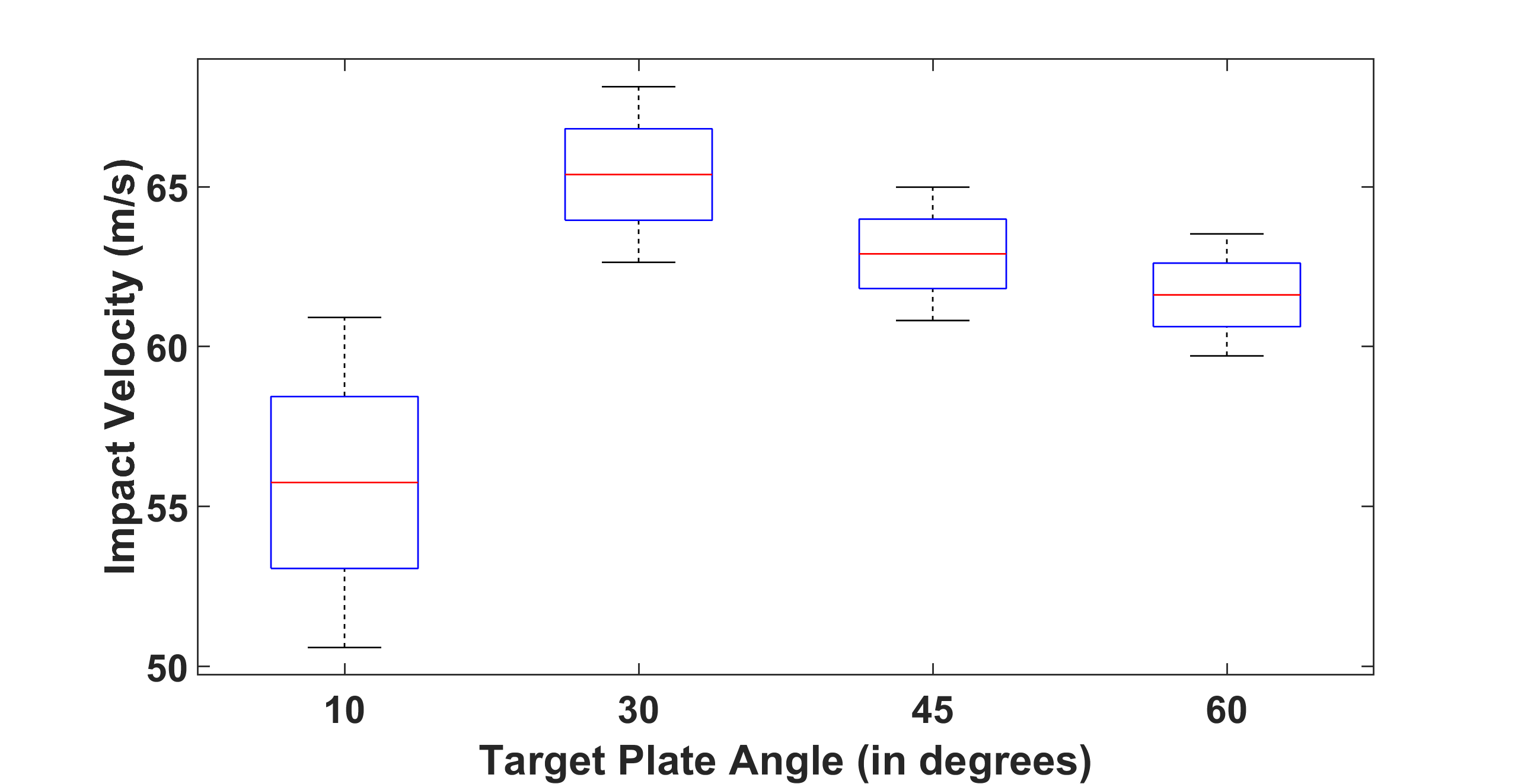}
		\caption{Target Plate Angle Test: Particle impact velocities (70 $\mu m$ diameter particles, $M_{exit} = 0.2$)}
		\label{fig:ang_vel} 
	\end{figure}

	The change in the angle of the target plate affects both the particle incidence angle as well as the impact velocity. Due to the high angle of attack of the target plate, a stagnation region, in \cref{fig:stag1} starts to form which in turn reduces the particle impact velocity.  \cref{fig:stag1} to \cref{fig:stag3} compare the effects of increasing the target angle of attack. At lower target plate angles, the particle impact locations increase in the axial direction. The boxplot in \cref{fig:ang_inc_ang} shows that on increasing the angle of attack of the plate, the particle incidence angle increases as expected. \cref{fig:ang_vel} shows that the particle impact velocities start to reduce at higher angles of attack. It is also interesting to note there seems to be larger variances in particle impact velocity as plate angle is decreased.
	
\begin{figure}[H]
		\includegraphics[width=0.9\textwidth]{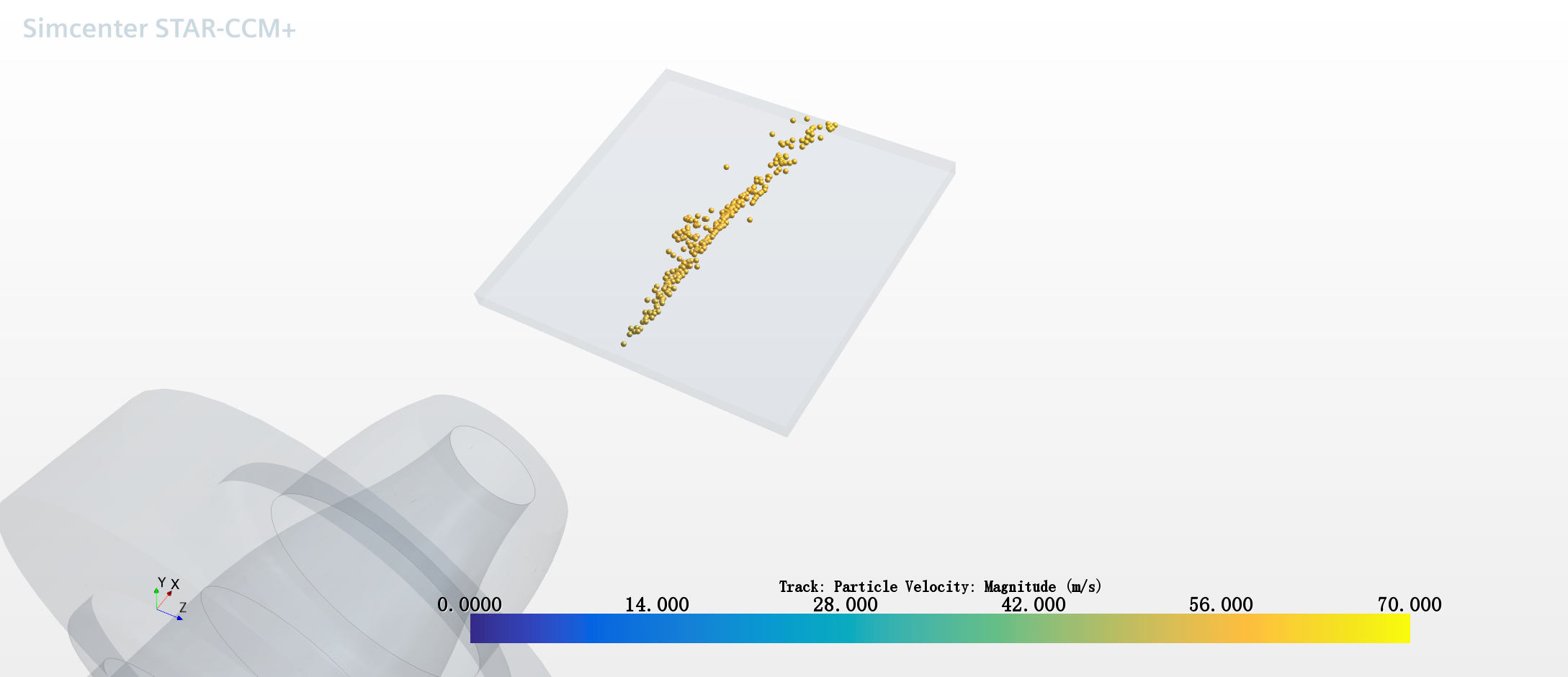}
		\caption{10$^\circ$ Particle impact ($M_{exit} = 0.2$)}
		\label{fig:10deg} 
	\end{figure}

\begin{figure}[H]
		\includegraphics[width=0.9\textwidth]{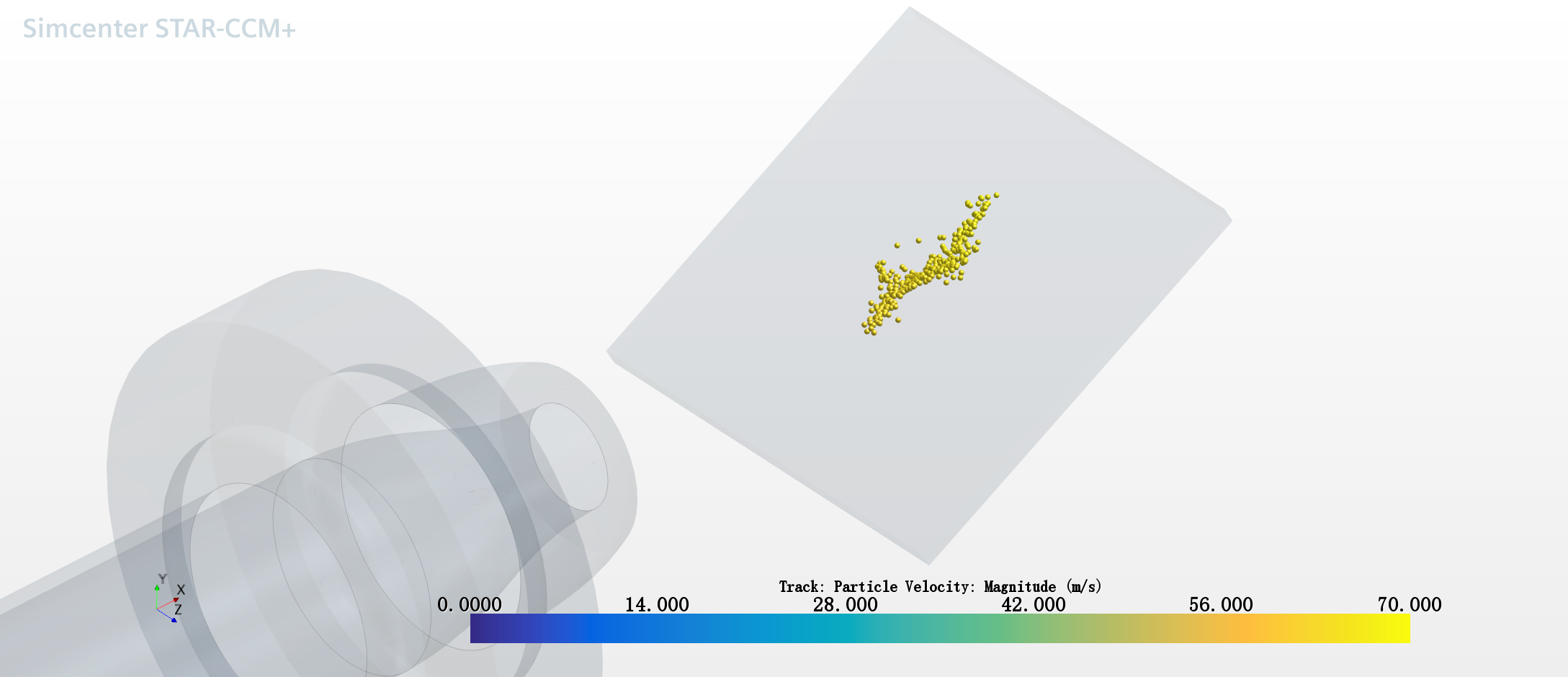}
		\caption{30$^\circ$ Particle impact ($M_{exit} = 0.2$)}
		\label{fig:30deg} 
	\end{figure}

\begin{figure}[H]
		\includegraphics[width=0.9\textwidth]{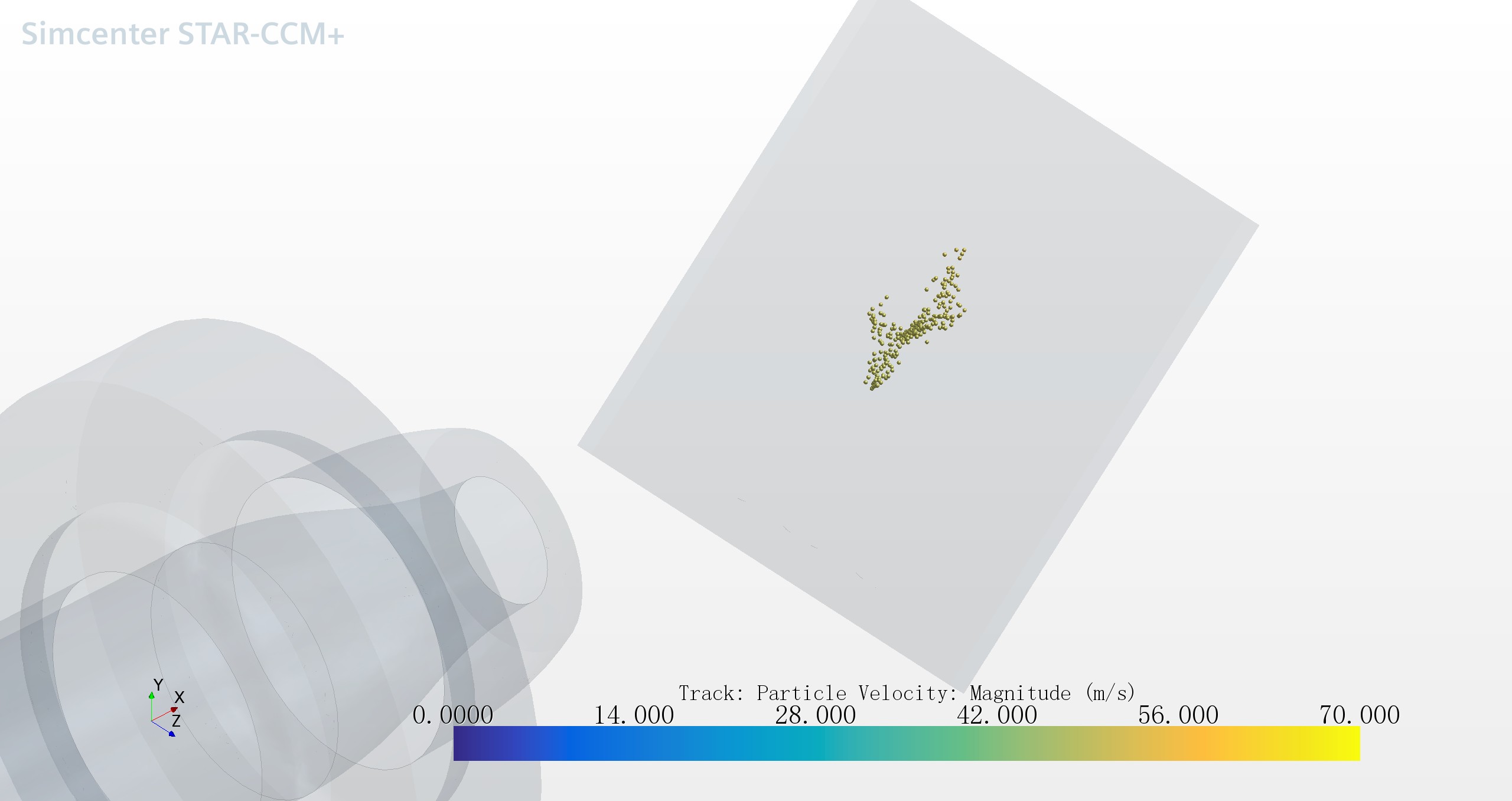}
		\caption{45$^\circ$ Particle impact ($M_{exit} = 0.2$)}
		\label{fig:45deg} 
	\end{figure}

\begin{figure}[H]
		\includegraphics[width=0.9\textwidth]{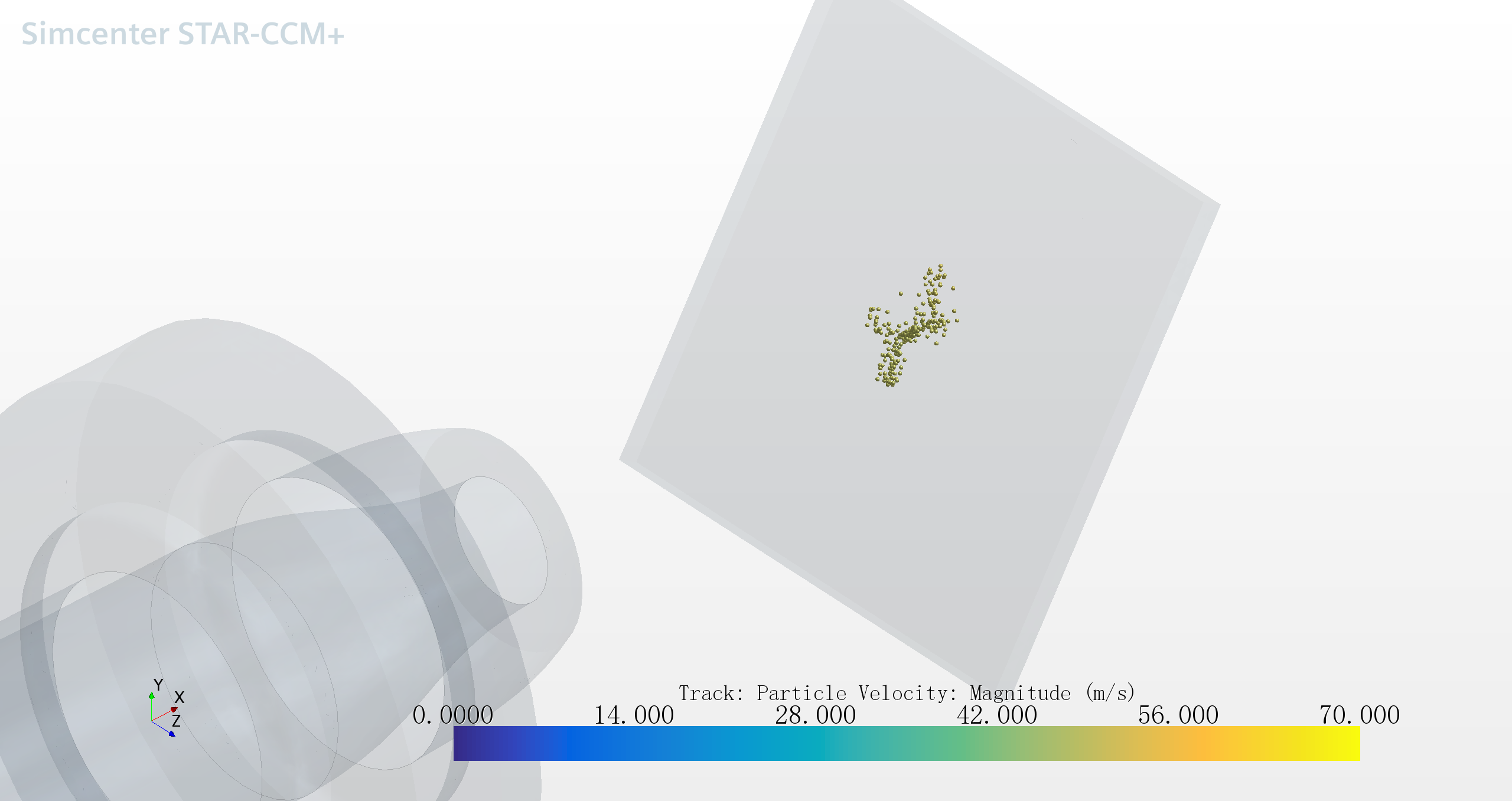}
		\caption{60$^\circ$ Particle impact ($M_{exit} = 0.2$)}
		\label{fig:60deg} 
	\end{figure}

	 \cref{fig:10deg} to \cref{fig:60deg} show the particle clustering on impact and we see that at 10$^\circ$ the particles tend to be more spread out in the axial direction which in turn also causes them to have a larger range of impact velocities, and as the plate angle of attack increases the particles have a more concentrated impact and so they also have a smaller range of impact velocities. This effect is likely purely a geometrical effect caused by the larger (smaller) projected area that the duct makes onto a low (high) angled target plate.
	
	\section{Comparisons To LES}\label{sec:LES}
	
	LES simulations are a useful tool for studying turbulence physics, as well as the development of turbulence models. Particle dispersion behavior is dependent on LES closure methods, the smaller particles are more susceptible to the dispersion and so a subgrid model for the particle motion can be effective \cite{Beck2019}. In StarCCM+, LES simulations do not implement any turbulent dispersion models. This is because in LES, the flow-field is decomposed into its resolved and modeled parts by filtering. StarCCM+ computes the resolved part of the turbulent dispersion, and since the dispersion of particles in LES is primarily due to the resolved scales, it is a reasonable assumption to ignore dispersion due to the modeled subgrid scales.
	
	The LES simulation implements the Wall-Adapting Local-Eddy Viscosity (WALE) subgrid scale model to solve for the subgrid scale turbulence. The WALE model is similar to the Smagorinsky subgrid scale model but is less sensitive to the value of the model coefficient. The WALE model implements an All $y^+$ wall treatment to calculate the wall shear stress and wall heat flux. The all-$y^+$ wall treatment uses blended wall functions and provides valid boundary conditions for flow, energy, and turbulence quantities for a wide range of near-wall mesh densities \cite{ccmSu}. 
	
	The Segregated Fluid Isothermal model in StarCCM+ maintains a constant temperature in the domain. Since the temperature variations are negligible in this study, it would be computationally expensive to solve the energy transport equation only for it to yield an almost constant temperature field, and so a Segregated Flow Solver is used for the LES simulations \cite{ccmSu}. The Segregated Flow Solver maintains the gas phase at a constant density as opposed to solving the ideal gas equation. A RANS simulation which uses the Segregated Flow model is used to obtain a steady flow solution which is then used as the initial conditions for the LES simulation to improve computational efficiency. The initial and boundary conditions are based on those used in the RANS simulations in Sec.\cref{sec:Setup}. The LES simulation uses a mass flow inlet for the gas phase which is calculated from the RANS simulations. The mass flow rate of the gas phase entering the domain is 0.1467686 $kg/s$. 70 $\mu m$ diameter particles are injected into the domain with the same injector conditions as shown in Tab.\cref{tab:properties}.
	
	In order to determine an appropriate time step for the LES simulation, it is necessary to choose a step that captures the vortex shedding in the duct. To do that, the Strouhal Number ($Sr$) for flow around the probes upstream of the particle seeder was calculated based on the studies by Fey et al. \cite{Fey1998}. The reason for choosing the probe diameter as the reference length is because it induces the fastest vortex shedding. For the simulations in this study, the Strouhal Number is 0.2096. The Strouhal Number is then used to calculate the vortex shedding frequency from the Eqn.\cref{eq:Sr} below, and the time step is the inverse of the shedding frequency.
	
	\begin{equation}
		Sr = f_{vortex}\frac{d_{duct}}{u_g}
		\label{eq:Sr}
	\end{equation}
	
	where, $f_{vortex}$ is the shedding frequency, $d_{duct}$ is diameter of the probes within the conveyor duct and $u_g$ is the gas velocity. From Eqn.\cref{eq:Sr}, the time step is approximately, 9e-4. With this time step the maximum convective Courant number was 30. In order to reduce the maximum convective Courant number to a value below 1, the time step was determined to be $3E-6 s$. This value ensures that the time step captures the vortex shedding from the smallest probe as well as maintains the maximum convective Courant number in the LES simulation to be less than 1. The mesh used for the simulations was the intermediate mesh which was discussed to be the optimal mesh size for the RANS simulations.
	
\begin{figure}[H]
		\centering \includegraphics[width=0.9\textwidth]{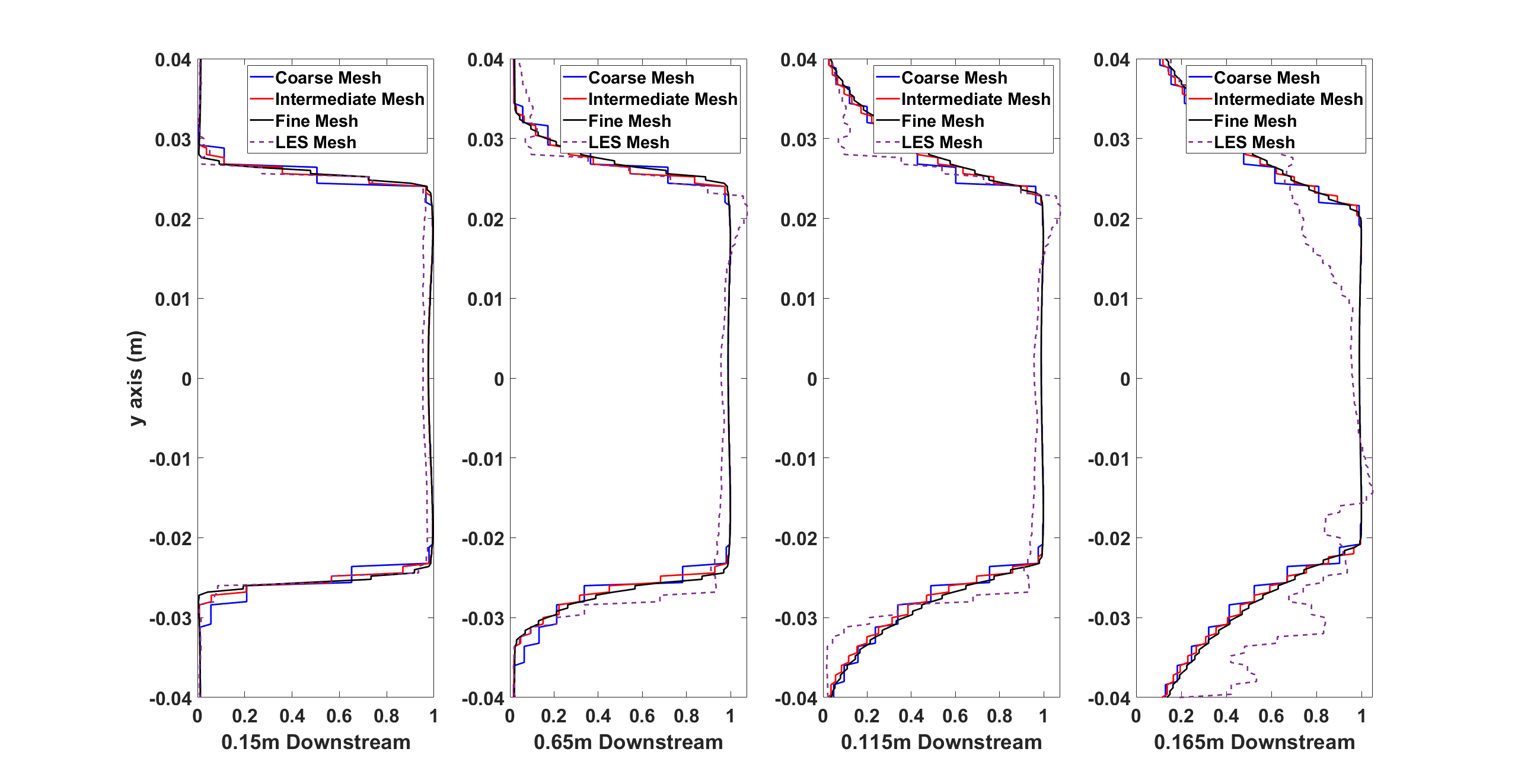}
		\caption{Gas Velocity at locations downstream of the nozzle  ($M_{exit} = 0.2$)}
		\label{fig:LES_gas_vel}
	\end{figure}

	The flow through time $t_{ft}$ is defined as the time taken by the gas phase to flow through the length of the duct, as was determined to be approximately 0.04s. The LES simulations were run for $2t_{ft}$ in order to achieve a fully developed flow that is statistically steady. Upon reaching statistical stationarity, the velocity profiles of the flow downstream of the nozzle are measured. The velocity profiles of the steady RANS meshes and the unsteady LES simulations are compared at multiple locations downstream of the nozzle exit. The unsteady nature of the LES simulations causes the profiles to vary slightly however the similarities between the LES and RANS simulations can be seen. \cref{fig:LES_gas_vel} shows the velocity profiles of the different mesh sizes used in the RANS grid study to the intermediate mesh size used in the LES study. As before, the actual grid data is demonstrated without interpolation, leading to a stair-step visualization artifact.
	
\begin{figure}[H]
		\centering \includegraphics[width=0.9\textwidth]{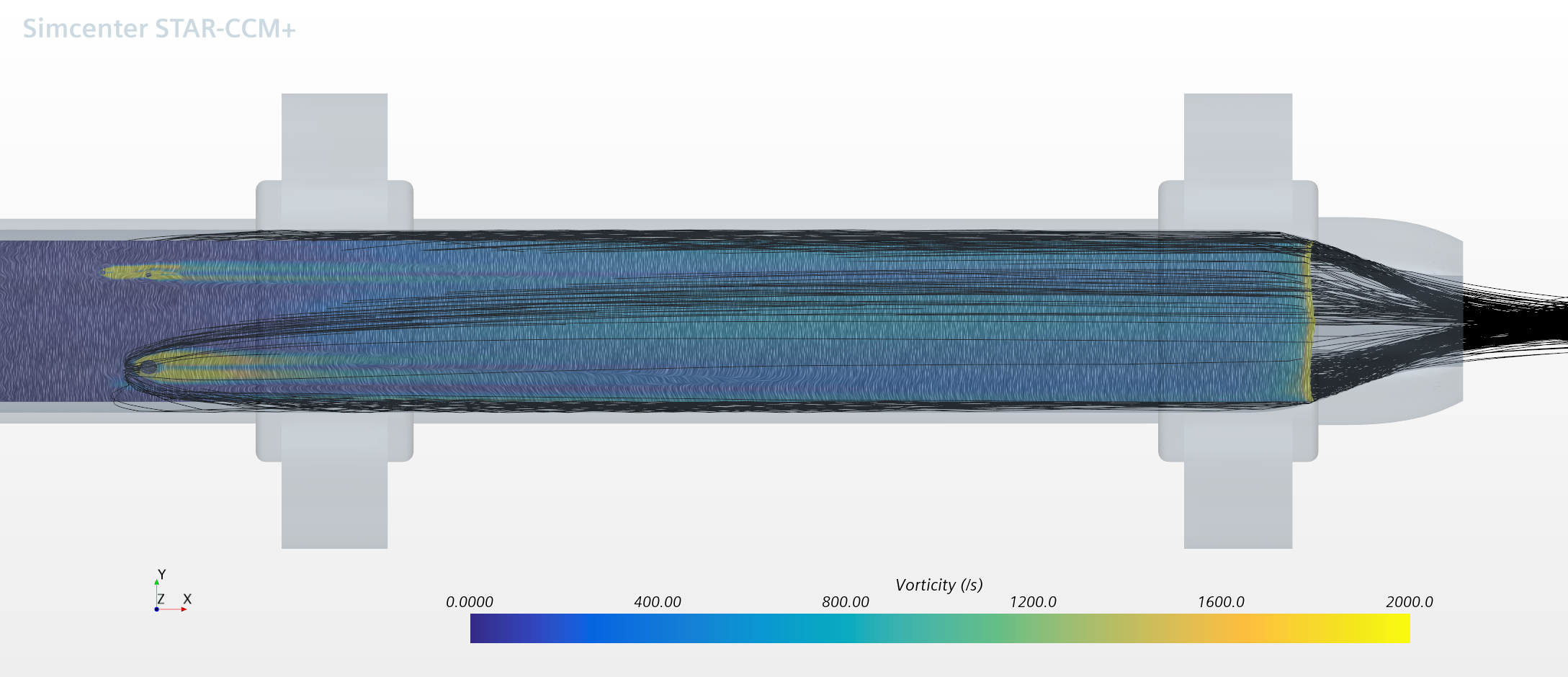}
		\caption{RANS simulation showing particle clustering \\ ($M_{exit} = 0.2$)}
		\label{fig:RANS_vortex}
	\end{figure}

\begin{figure}[H]
		\centering \includegraphics[width=0.9\textwidth]{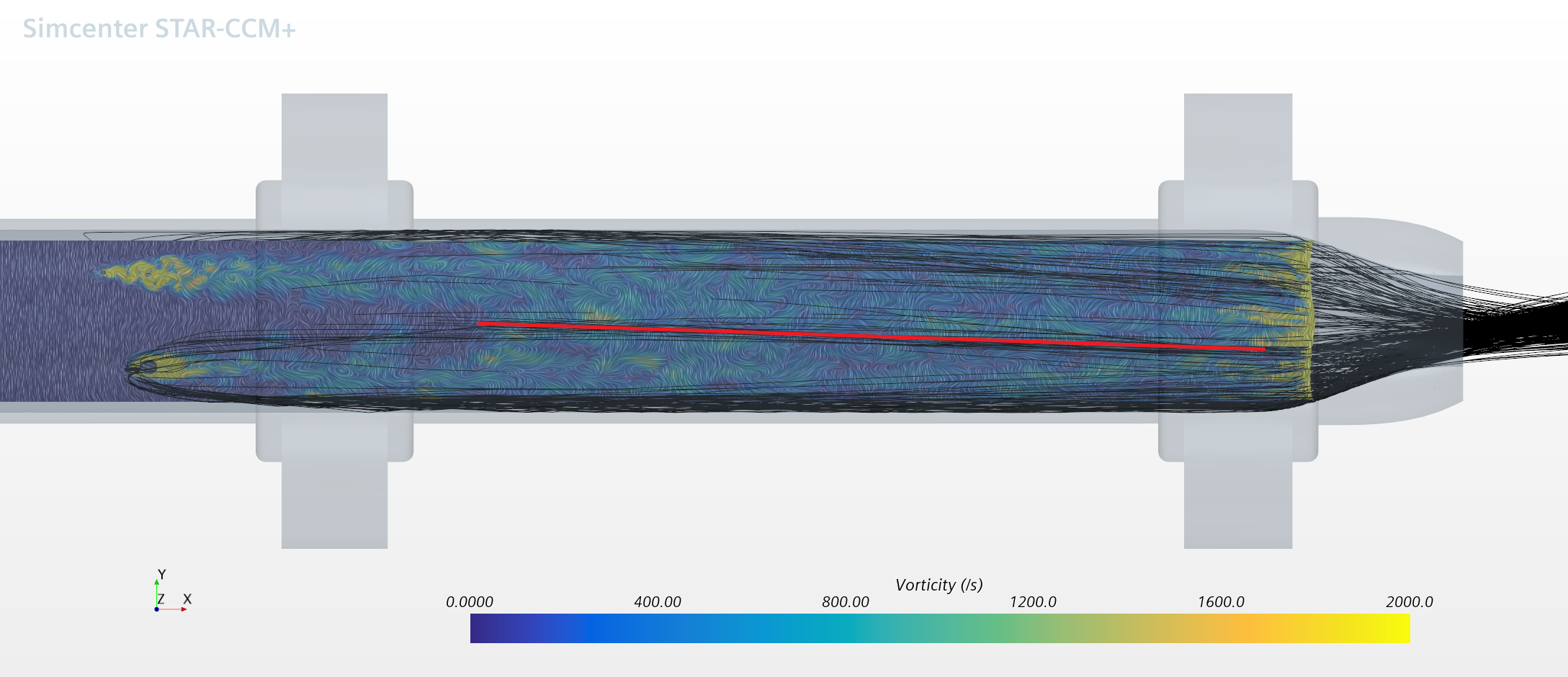}
		\caption{LES simulation showing particle clustering \\ ($M_{exit} = 0.2$)}
		\label{fig:LES_vortex}
	\end{figure}

	The pressure and temperature probes, along with the particle seeder tube generate vortices within the duct and the LES simulation is able to capture the vortex shedding within the duct with more accuracy. This difference in vortex shedding causes the flow profiles to differ between the RANS and LES simulations which in turn affects the particle trajectories within the duct. The difference in vortices formed are shown in \cref{fig:RANS_vortex} and \cref{fig:LES_vortex}. A cylindrical shell with a radius 93.5$\%$ of the conveyor duct radius is created to visualize the effects of the vortices on the particle tracks. This cylindrical contour is close enough to the duct surface without being affected by boundary layer effects. The recirculation region connected to the seeder tube are of similar lengths, and while the RANS shows a time averaged value, the LES simulation shows an instantaneous value of the vortex shedding. The vorticity magnitude contours are clipped to a maximum of 2000 /s to capture the shedding within the duct. 
	
	Due to the vortex shedding occurring because of the seeder tube, the flow starts to deviate from the axial direction, which in turn affects the particle path.  \cref{fig:RANS_vortex} and \cref{fig:LES_vortex} show the vorticity vectors on the same contour described above along with the particle tracks described by the black lines. The particle tracks are shown in black to contrast the vorticity contours.
	
	\cref{fig:RANS_vortex} and \cref{fig:LES_vortex} display the region where vortices formed due to the seeder tube and probe influence the particles in the RANS and LES simulations. The red line shows the merging of the two vortex shedding regions and the particle tracks follow the path where these two regions are merging, i.e the outer edges of both the vortex shedding regions. The particle Stokes Number in these simulations is 7.5 ($d_p = 70 \mu m$) and this is consistent with the theory that larger particles with $St \approx$ 1 tend to cluster around the periphery of vortices \cite{Miranda2020, Hwang2006}.
	
	However, these differences between the RANS and LES simulations of the flow within the duct do not affect the particle impact statistics. The particle impact velocities and impact angles are displayed in the box plot and compared to the RANS simulations. The particle trajectories near the target plate are also compared. There is a good comparison between the LES and RANS for impact velocity while the LES has a larger range of impact angles but the mean is still similar to RANS simulations. This seems to indicate that the RANS simulation data can correctly capture many aspects of the physics of this problem.
	
	\begin{figure}[H]
		\centering \includegraphics[width=0.9\textwidth]{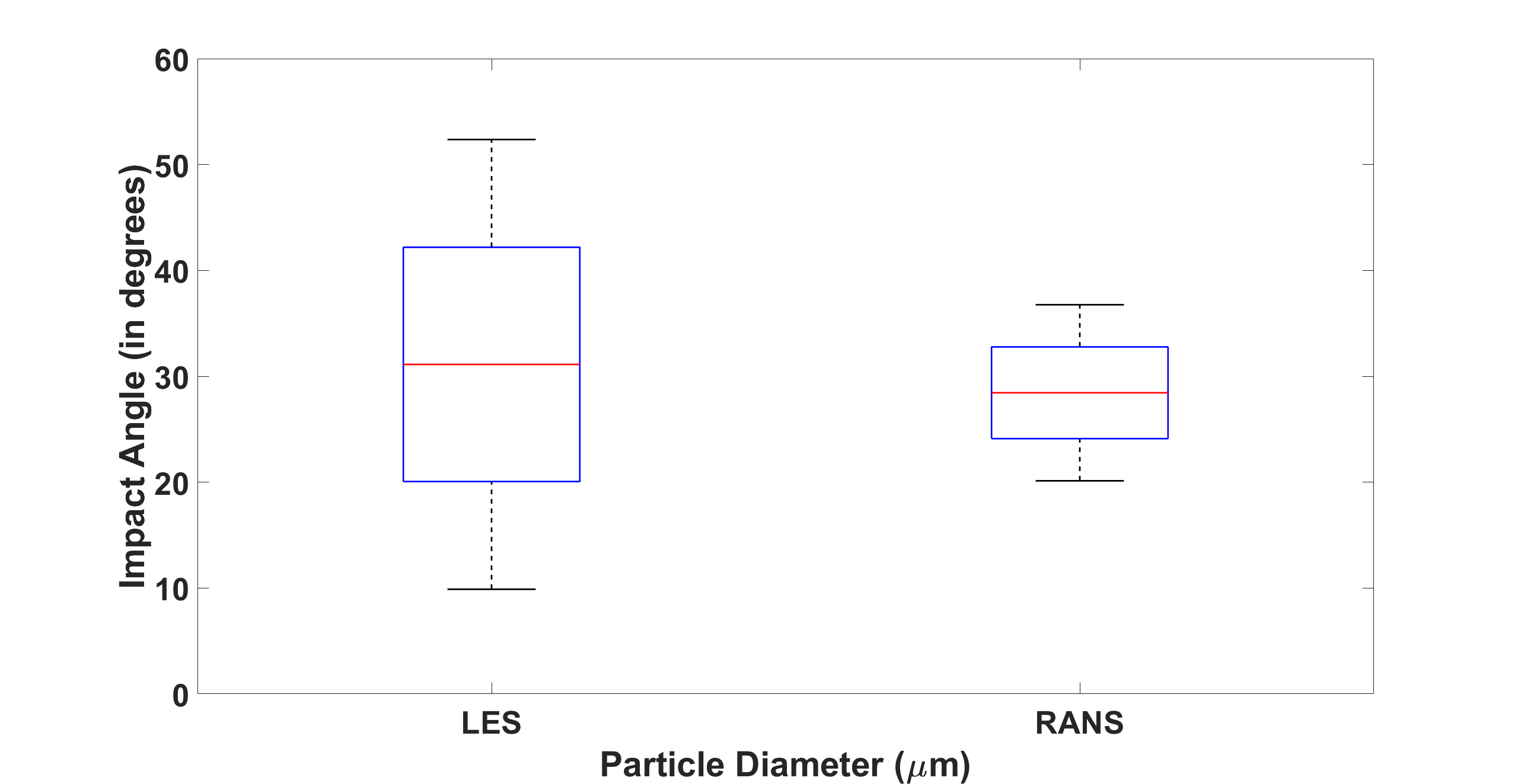}
		\caption{Target Plate Test: Particle incident angles ($M_{exit} = 0.2$)}
		\label{fig:LES_vel} 
	\end{figure}
	
	\begin{figure}[H]
		\centering \includegraphics[width=0.9\textwidth]{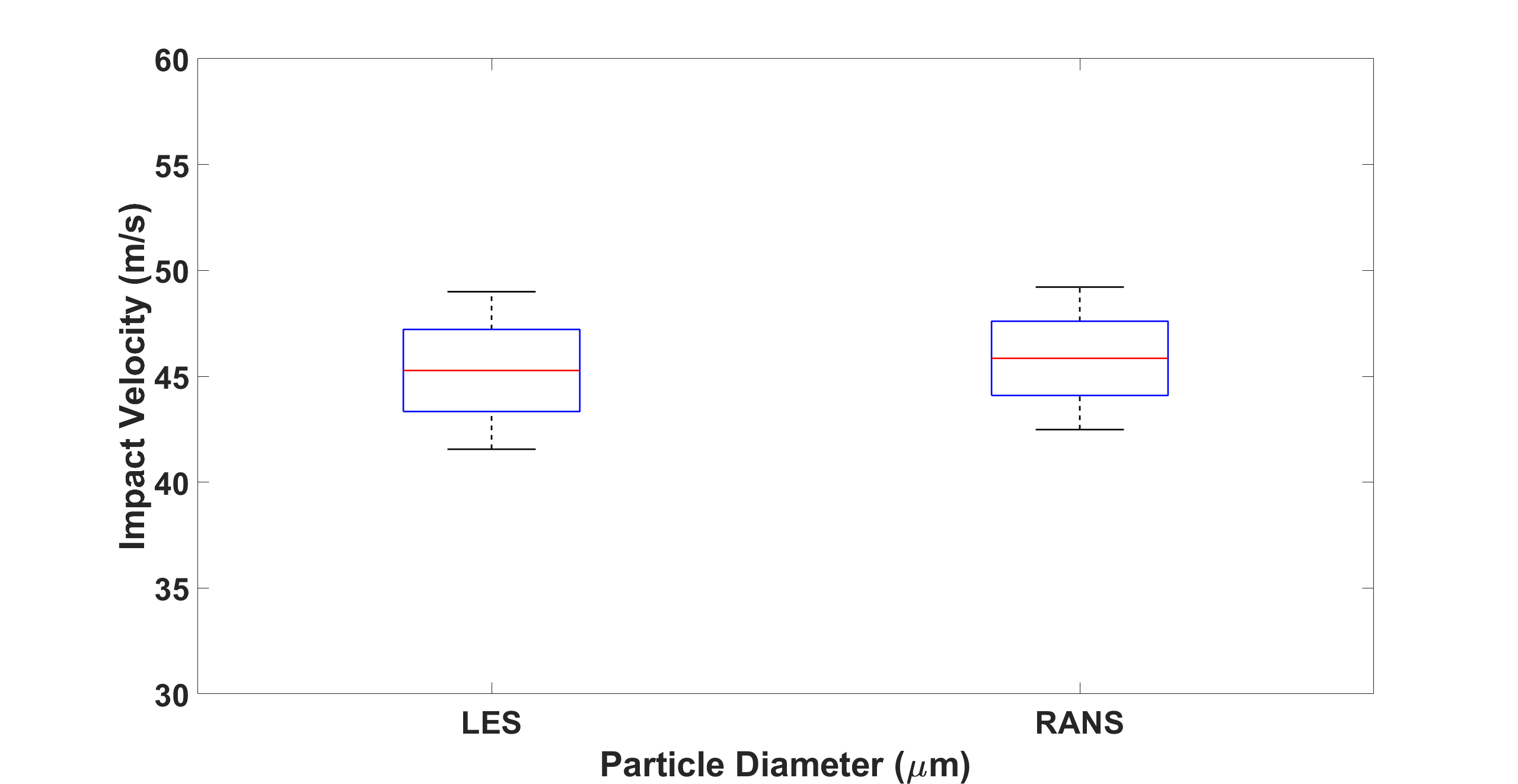}
		\caption{Target Plate Test: Particle impact velocities ($M_{exit} = 0.2$)}
		\label{fig:LES_inc} 
	\end{figure}

	\section{Conclusion}
	The LPT-RANS studies were able to predict the particle impact speeds, incidence angles and the particle speeds at the nozzle exit. They also helped understand the particle impact locations based on particle size. The target plate impact velocity depends on both particle exit velocity and inertia (i.e. particle size), so measuring exit velocity alone is insufficient for experiments. For the same reason, particle incidence angles are also strongly dependent on particle size. 
	
	The particle trajectories show the adverse effects of the near wall aerodynamic effects just before collision. The particle velocities are influenced significantly by these effects, and in order to predict particle rebound properties and erosion accurately, these effects will have to be taken into account.
	
	The LES simulations show good comparison to the RANS simulations, when the statistical particle impact data is compared as well as the velocity profiles downstream of the nozzle exit. The vortices within the duct are formed due to vortex shedding behind the probes and particle seeder which in turn cause the particle trajectories inside the conveyor tube to be different between the RANS and LES simulations. This is because particles with diameter, 70 $\mu m$ have a Stokes Number of around 7, which causes them to be affected by the vortices formed in the duct. As the LES simulations capture the vortices more accurately, they affect the particle trajectories more. Although the particle trajectories within the duct are different between steady and unsteady, the statistical information at the target plate seems to be consistent between the two.

	\section*{Acknowledgments}
	The research was performed as part of the Rolls-Royce University Technology Centre a Virginia Tech. C. Miranda was funded as a Virginia Tech Rolls-Royce Doctoral Fellow. The authors thank Mr. Jim Loebig (Rolls-Royce plc) for providing technical expertise and research guidance for this work. The authors acknowledge Advanced Research Computing at Virginia Tech for providing computational resources and technical support that have contributed to the results reported within this paper.

	\bibliographystyle{asmejour} 
	
	\bibliography{asmejour-sample}

\end{document}